{}
{}

\def\verdraft{3}
\def\verpublish{4}
\def\verarxiv{5}

\def\version{5}

\documentclass[12pt,a4paper]{article}

\usepackage{amsmath,amssymb}
\ifnum\version<2\usepackage{showkeys}\fi
\usepackage{graphicx}
\usepackage[bookmarks=true,hyperfigures=true]{hyperref}
\usepackage[usenames,dvipsnames]{color}
\usepackage[nosort]{cite}
\usepackage[bulletsep]{collref}
\usepackage[utf8]{inputenc}

\def\showkeyslabelformat#1{{\usefont\encodingdefault\familydefault%
 \seriesdefault\shapedefault\normalfont\tiny\ttfamily#1}}
\def\showkeysrefformat#1{{\normalfont\tiny\ttfamily#1}}
\makeatletter
\def\SK@@ref#1>#2\SK@{%
 {\@inlabelfalse\leavevmode\vbox to\z@{%
 \vss\SK@refcolor\rlap{\vrule\raise .75em%
  \hbox{\showkeysrefformat{#2}}}}}}
\makeatother

\usepackage[a4paper,text={160mm,247mm},centering]{geometry}

\RequirePackage{verbatim}

\makeatletter
\ifnum\version<\verarxiv\relax
\RequirePackage{shellesc}
\ifnum\ShellEscapeStatus>0\relax
\AddToHook{enddocument/afteraux}{\ShellEscape{bibtex \jobname}}
\fi
\newwrite\bibinl@out
\newenvironment{bibtex}[1][\jobname]{%
  \immediate\openout\bibinl@out #1.bib
  \immediate\write\bibinl@out{\@percentchar generated from `\jobname' starting line \the\inputlineno^^J}%
  \def\verbatim@processline{\immediate\write\bibinl@out{\the\verbatim@line}}%
  \@bsphack\let\do\@makeother\dospecials\catcode`\^^M\active\verbatim@start
}%
{\immediate\closeout\bibinl@out\@esphack}
\else
\newenvironment{bibtex}[1][\jobname]{\comment}{\endcomment}
\fi
\makeatother

\counterwithin{equation}{section}

\usepackage[font=small,labelfont=bf,width=0.85\textwidth]{caption}

\usepackage{eqnlines}
\eqnlinesset{number=on,numberline=center}
\eqnlinesset{punctall={,,.}}

\ifnum\version<2
\PassOptionsToPackage{now,nowall}{mpostinl}
\fi
\ifnum\version=\verarxiv
\PassOptionsToPackage{write=false,compile=false}{mpostinl}
\fi

\RequirePackage{graphbox}
\RequirePackage[clean=false]{mpostinl}

\begin{mposttex}
\usepackage{amsmath,amssymb}
\usepackage[utf8]{inputenc}
\usepackage[usenames, dvipsnames]{color}
\end{mposttex}

\begin{mpostdef}
def pensize(expr s)=withpen pencircle scaled s enddef;
def fillshape(expr p,c)=
  fill p withcolor c;
  draw p
enddef;
def drawdot(expr z,s)=
  fill fullcircle scaled s shifted z
enddef;
def filldot(expr z,s,ci)=
  fillshape(fullcircle scaled s shifted z, ci)
enddef;
def drawcross(expr z,s,r,t,c)=
  draw ((-0.5,-0.5)--(+0.5,+0.5)) scaled s rotated r shifted z pensize(t) withcolor c;
  draw ((+0.5,-0.5)--(-0.5,+0.5)) scaled s rotated r shifted z pensize(t) withcolor c;
enddef;
\end{mpostdef}

\begin{mpostdef}
pair vpos[];
numeric nums[];
path paths[];
newinternal numeric xu;
xu:=1cm;
ahlength:=4pt;
\end{mpostdef}

\begin{mpostdef}
def midarrowperc (expr p, t) =
  fill arrowhead subpath(0,arctime(t*arclength(p)+0.5ahlength) of p) of p;
enddef;
def midarrow (expr p, t) =
  fill arrowhead subpath(0,arctime(arclength(subpath (0,t) of p)+0.5ahlength) of p) of p
enddef;
\end{mpostdef}

\begin{mpostdef}
def rbox (expr p,s,r) =
  (((r,0)--(xpart(s)-r,0){dir 0}..{dir 90}
   (xpart(s),r)--(xpart(s),ypart(s)-r){dir 90}..{dir 180}
   (xpart(s)-r,ypart(s))--(r,ypart(s)){dir 180}..{dir 270}
   (0,ypart(s)-r)--(0,r){dir 270}..{dir 0}
  cycle) shifted (p-0.5s))
enddef;
\end{mpostdef}

\begin{mposttex}
\def\version{1}
\end{mposttex}
\ifnum\version=11
\begin{mposttex}
\def\version{11}
\end{mposttex}
\fi

\begin{mposttex}[dual]

\ifnum\version=11
\usepackage[bitstream-charter]{mathdesign}
\fi

\usepackage{mathfixs}
\ProvideMathFix{autobold,greekcaps,frac,root}
\ProvideMathFix{multskip}
\ProvideMathFix{numsets,Order*,res}

\ProvideMathFix{der,diff,hyp,tr,diag,reim,mapclass,span,defeq,eqdef,ad,id}
\newcommand{\wanteq}{\stackrel{!}{=}}

\RequirePackage[extdef]{delimset}

\ProvideMathFix{vfrac,rfrac,iunit*nb,econst,halves}

\newcommand{\alg}[1]{\mathfrak{#1}}
\newcommand{\grp}[1]{\mathrm{#1}}
\newcommand{\gen}[1]{\mathrm{#1}{}}

\newcommand{\permop}{\operatorname{ex}}

\newcommand{\spc}[1]{\mathbb{#1}}

\newcommand{\cybe}{\pair[S,]}
\newcommand{\liebr}{\comm}
\newcommand{\cobra}{\delta}

\newcommand{\rmat}{r}
\newcommand{\Rmat}{R}
\newcommand{\kmat}{k}
\newcommand{\Kmat}{K}
\newcommand{\urmat}{\mathrm{\rmat}}

\newcommand{\ukmat}{\mathrm{\kmat}}

\newcommand{\ctrig}{{\text{trig}}}
\newcommand{\dirrefl}{n}
\newcommand{\nullrefl}{\eta}
\newcommand{\dirreflL}{\dirrefl}
\newcommand{\dirreflP}{\dirrefl_\labP}
\newcommand{\dirtw}{\xi}
\newcommand{\ptfix}{{0}}
\newcommand{\ptrefl}[1][]{{\text{r#1}}}
\newcommand{\parrefl}{w}
\newcommand{\parreflL}{\parrefl}
\newcommand{\parreflP}{\parrefl_\labP}
\newcommand{\parfix}{u_{\ptfix}}
\newcommand{\parfixP}{u_{\ptfix,\labP}}
\newcommand{\parrot}{\rho}

\newcommand{\slstr}{f}
\newcommand{\slinv}{c}
\newcommand{\slpauli}{\tilde\sigma}
\newcommand{\sleps}{\varepsilon}
\newcommand{\slgen}{\gen{J}}

\newcommand{\sldot}{{\cdot}}

\newcommand{\labA}{{\text{A}}}
\newcommand{\labB}{{\text{B}}}
\newcommand{\labAB}{{\text{A/B}}}
\newcommand{\labBA}{{\text{B/A}}}
\newcommand{\labL}{{\genL}}
\newcommand{\labP}{{\genP}}
\newcommand{\labM}{{\genM}}
\newcommand{\labH}{{\genH}}
\newcommand{\labK}{{\genK}}

\newcommand{\genJ}{\gen{J}}
\newcommand{\genL}{\gen{L}}
\newcommand{\genP}{\gen{P}}
\newcommand{\genM}{\gen{M}}
\newcommand{\genD}{\gen{D}}
\newcommand{\genDL}{\genD_\labL}
\newcommand{\genDP}{\genD_\labP}
\newcommand{\genC}{\gen{C}}
\newcommand{\genQ}{\gen{Q}}
\newcommand{\genH}{\gen{H}}
\newcommand{\genK}{\gen{K}}
\newcommand{\genLt}{\genL}
\newcommand{\genPt}{\genP}
\newcommand{\genDLt}{\tilde{\genD}_\labL}

\newcommand{\casJ}{\genJ^2}
\newcommand{\casQ}{\genQ^2}

\newcommand{\mref}{{\bar m}}
\newcommand{\mratio}{{\mu}}

\end{mposttex}

\providecommand{\href}[2]{#2}

\makeatletter
\def\mr@ignsp#1 {\ifx\:#1\@empty\else #1\expandafter\mr@ignsp\fi}%
\newcommand{\multiref}[1]{\begingroup
\xdef\mr@no@sparg{\expandafter\mr@ignsp#1 \: }%
\def\mr@comma{}%
\@for\mr@refs:=\mr@no@sparg\do{\mr@comma\def\mr@comma{,}\ref{\mr@refs}}%
\endgroup}
\renewcommand{\eqref}[1]{(\multiref{#1})}
\makeatother

\makeatletter
\newcommand{\namedref}[2]{\hyperref[#2]{#1~\ref*{#2}}}
\newcommand{\secref}{\@ifstar{\namedref{Section}}{\namedref{Sec.}}}
\newcommand{\appref}{\@ifstar{\namedref{Appendix}}{\namedref{App.}}}
\newcommand{\tabref}{\@ifstar{\namedref{Table}}{\namedref{Tab.}}}
\newcommand{\figref}{\@ifstar{\namedref{Figure}}{\namedref{Fig.}}}
\makeatother

\edef\thebibliography{\unexpanded{\phantomsection\addcontentsline{toc}{section}{\refname}}
  \unexpanded\expandafter{\thebibliography}}
\edef\tableofcontents{\unexpanded{\phantomsection\addcontentsline{toc}{section}{\contentsname}}
  \unexpanded\expandafter{\tableofcontents}}

\providecommand{\hypersetup}[1]{}
\providecommand{\texorpdfstring}[2]{#1}

\hypersetup{plainpages=false}
\hypersetup{pdfpagemode=UseNone}
\hypersetup{bookmarksnumbered=true}
\hypersetup{pdfstartview=FitH}
\hypersetup{colorlinks=false}
\hypersetup{citebordercolor={.5 1 .5}}
\hypersetup{urlbordercolor={.5 1 1}}
\hypersetup{linkbordercolor={1 .7 .7}}

\ifnum\version<\verpublish\PassOptionsToPackage{draft}{metastr}\fi
\RequirePackage{metastr}

\ifnum\version<\verdraft\metasetterm{draft}{EDITING}\fi

\usepackage{ifpdf}
\ifpdf\else\RequirePackage[active]{srcltx}\fi
\RequirePackage{color}

\newcommand{\remark}[2][]{}

\newcommand{\hlcolor}{\color}
\ifnum\version<\verpublish\else
\renewcommand{\hlcolor}[2][]{}
\fi

\ifnum\version<\verpublish
\renewcommand{\remark}[2][]{\ifmmode\hbox\fi{%
  \def\emph{\textsl}\def\textbullet{$\bullet$}%
  \def\tmparga{#1}\def\tmpargb{draft}\ifx\tmparga\tmpargb%
  \color[rgb]{0.5,0,0}\normalfont\sffamily\hspace{1ex} #2\hspace{1ex}\fi}}
\fi

\ifnum\version<\verdraft
\renewcommand{\remark}[2][]{\ifmmode\hbox\fi{\normalfont\sffamily\hspace{1ex}%
  \def\emph{\textsl}\def\textbullet{$\bullet$}%
  \def\tmparga{#1}%
  \def\tmpargb{NB}\ifx\tmparga\tmpargb\color[rgb]{0,0,0.6}\fi%
  \def\tmpargb{DL}\ifx\tmparga\tmpargb\color[rgb]{0,0.6,0}\fi%
  \def\tmpargb{new}\ifx\tmparga\tmpargb\color[rgb]{0.6,0,0}\fi%
  \def\tmpargb{old}\ifx\tmparga\tmpargb\color[rgb]{0.5,0.5,0.5}\fi%
  \def\tmpargb{}\ifx\tmparga\tmpargb\color{red}\fi%
  \def\tmpargb{}\ifx\tmparga\tmpargb\else \textbf{#1:} \fi%
  #2\hspace{1ex}}}
\fi

\metaset{title}{Reflection Matrices\texorpdfstring{\\}{ }by\texorpdfstring{\\}{ }Contraction and Reduction}
\metaset{author}{Niklas Beisert, Danilo Lučić}

\begin{document}

\ifnum\version>0
\pdfbookmark[1]{Title Page}{title}
\thispagestyle{empty}

\begingroup\raggedleft\footnotesize\ttfamily
\par\endgroup

\vspace*{2cm}
\begin{center}%
\begingroup\Large\bfseries\metapick[print]{title}\par\endgroup
\vspace{1cm}

\begingroup\scshape
\metapick[print]{author}
\endgroup
\vspace{5mm}

\textit{Institut für Theoretische Physik,\\
Eidgenössische Technische Hochschule Zürich,\\
Wolfgang-Pauli-Strasse 27, 8093 Zürich, Switzerland}
\vspace{0.1cm}

\begingroup\ttfamily\small
nbeisert@itp.phys.ethz.ch\\
d.lucic@studenti.unipi.it\par
\endgroup
\vspace{5mm}

\vfill

\metaif[]{draft}{
\vspace{5mm}
\begingroup\bfseries\Large \metaterm{draft}\endgroup
\par
\vspace{5mm}
\vfill
}{}

\textbf{Abstract}\vspace{5mm}

\begin{minipage}{12.7cm}
We reconsider the construction of reflection matrices
for the factorised AdS/CFT worldsheet scattering problem
and for the integrable one-dimensional Hubbard model.
These systems are governed by an unconventional 
quasi-triangular superalgebra based on $\alg{psu}(2|2)$
which was shown in previous work to derive from
the exceptional simple superalgebra $\alg{d}(2,1;\epsilon)$ at small $\epsilon$ 
via the method of contraction and reduction.
In this work, we apply contraction and reduction 
to derive reflection matrices for the above systems in the classical limit
from elementary ones. 
We show that the method is compatible
with an abstract algebraic notion of k-matrices
as group-like objects implementing reflection automorphisms.
We furthermore investigate the space of allowable parameters 
and find several inequivalent families of k-matrices
in agreement with previous constructions in the AdS/CFT context.
\end{minipage}

\vspace*{4cm}

\end{center}

\ifnum\version>1
\newpage
\tableofcontents
\fi

\newpage
\fi

\section{Introduction}
\label{sec:intro}

The study of factorised scattering on the worldsheet
in the AdS/CFT-context \cite{Staudacher:2004tk} (see \cite{Beisert:2010jr} for reviews)
has brought about a peculiar two-particle scattering matrix \cite{Beisert:2005tm}.
It was subsequently shown in \cite{Beisert:2006qh} that
this particular matrix structure had been constructed much earlier
as the quantum R-matrix for the integrable one-dimensional Hubbard model \cite{Shastry:1986bb}
(see \cite{Essler:2005aa} for a review).
The matrix has a number of unusual features.
Most evidently, it has a rich functional and combinatorial structure,
and the matrix elements do not follow a difference form
as is commonly the case for R-matrices satisfying the 
Yang--Baxter equation. As such, it was initially unclear
whether and how this R-matrix fits the established scheme
for quantum algebras based on semi-simple Lie algebras and superalgebras.

Aspects of a suitable quantum algebra were developed in 
\cite{Gomez:2006va,Plefka:2006ze,Beisert:2006fmy},
but a formulation of the complete quantum algebra and its universal R-matrix
seemed out of reach due to their inherent complexity.
The problem was considerably simplified by considering a classical limit
where most non-linear expressions and relations of the quantum algebra linearise.
The classical limit yielded a quasi-triangular Lie bialgebra with a classical r-matrix
\cite{Klose:2006zd,Torrielli:2007mc,Moriyama:2007jt,Beisert:2007ty}.
The resulting loop algebra turned out to be based on
a non-simple extension of the Lie superalgebra $\alg{psu}(2|2)$.
Moreover, the loop algebra extends to an affine algebra 
whose affine derivation has a non-trivial cobracket \cite{Beisert:2010kk}
which ultimately explains the non-difference form of the r-matrix.

The work \cite{Beisert:2017xqx} initiated a new approach for construction of the R-matrix
based on the exceptional simple superalgebra $\alg{d}(2,1;\epsilon)$ with a continuous parameter $\epsilon$.
For $\epsilon=0$ this superalgebra reduces to $\alg{psu}(2|2)$,
and by supplementing an $\alg{sl}(2)$ algebra
one can define a contraction limit $\epsilon\to0$
yielding the extension $\alg{sl}(2)\ltimes\alg{psu}(2|2)\ltimes\Real^{2,1}$
which is relevant to the above r-matrix.
The affine algebra thereof was subsequently reduced in \cite{Beisert:2022vnc,Beisert:2024rke} 
to a deformed affine algebra of $\alg{gl}(1)\ltimes\alg{psu}(2|2)\ltimes\Real$
serving as the symmetry algebra of the r-matrix.
This method of contraction and reduction thus yields 
a non-simple algebra from a semi-simple one, and importantly,
it preserves the quasi-triangular structure that extends the algebra
with an r-matrix satisfying the classical Yang--Baxter equation.
In other words, it opens a path for deriving the above r-matrix 
from a standard r-matrix associated to a semi-simple,
yet exceptional, Lie superalgebra.
Moreover, the method works at the level of the abstract algebraic relations,
and it therefore reproduces a universal, i.e.\ representation-independent, classical r-matrix.
When the contraction and reduction method is lifted to the level of quantum algebras \cite{Beisert:2026xxx}, 
one can expect to obtain the corresponding universal quantum R-matrix in this way.

In the present work, we extend the method of contraction and reduction 
to the notion of reflection matrices which are also known as K-matrices
in the context of quantum algebra.
Within the scattering framework on the AdS/CFT worldsheet,
several classes of reflection matrices have been constructed
\cite{Berenstein:2005vf,Hofman:2007xp,Ahn:2007bq,Ahn:2008df,Murgan:2008zu,Correa:2008av,Correa:2011nz},
investigated towards their symmetry algebra \cite{Ahn:2010xa,MacKay:2010ey,MacKay:2011av,Palla:2011eu,MacKay:2011zs,deLeeuw:2012fy},
and applied towards spectral analysis \cite{Galleas:2009ye,Nepomechie:2009zi,Bajnok:2012xc},
see \cite{Zoubos:2010kh,Ciavarella:2011jca,Regelskis:2012voq} for reviews. 
Each of these reflection matrices implements some non-trivial open boundary condition imposed by 
the string worldsheet ending on some D-brane.
A worldsheet excitation meeting such an obstacle will be reflected
with a specific phase (and potential change of flavour) which is described
by the reflection matrix.
Within a factorised scattering setup, reflection matrices factorise
into a sequence of single-particle reflection and two-particle scattering processes.
Next to the bulk Yang--Baxter equation, consistent factorisation
requires the reflection equation,
which is also known as the boundary Yang--Baxter equation
\cite{Cherednik:1984vvp,Sklyanin:1988yz,Kulish:1992qb}.

Our aim is to show how to derive the AdS/CFT reflection matrices
from reflection matrices for elementary semi-simple Lie (super)algebras.
However, reflection matrices are usually expressed for concrete representations.
The contraction and reduction method also applies at the level of representations,
but especially the contraction step is somewhat challenging, see \cite{Beisert:2022vnc}.
Here, we prefer to work at the level of abstract algebra,
and we therefore need an abstract algebraic notion for the reflection matrix,
a universal K-matrix, and its classical counterpart which we call the (classical) k-matrix.
In this way we can avoid the specialisation to specific representations
and make general statements which apply to arbitrary representations.
In this framework, a k-matrix is the abstraction of a specific automorphism of the loop algebra.
This automorphism is a large, group-like transformation whereas the elements
of the loop algebra define infinitesimal, algebra-like transformations. As such,
the action of the k-matrix on the loop algebra is defined via its adjoint action.
The classical boundary Yang--Baxter equation is then expressed
through the k-matrix adjoint actions on the r-matrix.
Equipped with a suitable k-matrix for the starting semi-simple Lie algebra,
we can apply the contraction and reduction procedure to
obtain a k-matrix for the resulting non-simple algebra.

One significant obstacle in this construction,
particularly towards its presentation, is complexity. 
The full-fledged construction uses various concepts
of classical or quantum algebra, 
different amounts of supersymmetry, 
particular Drinfeld twists and potentially q-deformation.
Furthermore, it starts with non-trivial parameter spaces of reflection matrices
but ends with several inequivalent discrete instances of them.
The construction and its several steps
work out with many, but not all, choices for the ingredients.
Evidently, we aim towards the complete setup
which will lead to the reflection matrices encountered in the AdS/CFT context.
However, the full treatment obscures the method,
and we prefer to expose the steps of the construction using a basic setup instead.
We will then add some of the more advanced ingredients step by step.
While the construction steps will largely remain the same, 
what changes significantly for each step and each ingredient,
is the space of permissible parameter values of the reflection matrices.
Some construction steps preserve the parameter space,
others constrain and separate it into different cases which require individual discussion;
and some steps can even enhance the parameter space. 
As we do not know, a priori, at which point in parameter space we have to start
in order for the construction steps to pass,
we will explore and expose the parameter space at the individual steps of the construction
and at different levels of the complexity.
The figures in \eqref{eq:fig-untwisted} and \eqref{eq:fig-twisted}
can serve as a guide to the flow of our deductions.
Despite our efforts, whose results are not always clean and pretty,
this treatment is not meant to be exhaustive.
It is rather meant to demonstrate that there are suitable starting points
and to guide us in setting them up
such that the construction will pass and lead to the desired reflection matrices.

This work is structured as follows:
In \secref{sec:kmat}, we introduce the framework of scattering 
and reflection matrices for factorised scattering.
We will also expose the elementary expressions
for such matrices for the simplest algebra $\alg{sl}(2)$
and introduce a specific class of Drinfeld twists
that will be needed for the reduction step later on.
In \secref{sec:contractreduct}, we explain the contraction and reduction method
and apply it to reflection matrices.
In particular, we will investigate how the set of parameters
for the reflection matrices changes at each step of the construction. 
Here, we will arrive at several distinct choices of reflection matrices
with some residual continuous parameters.
\secref{sec:supersymmetry} enhances the construction 
by supersymmetry and yields some extra families of reflection matrices.
We conclude in \secref{sec:conclusions} and give an outlook.

\section{Reflection Matrices}
\label{sec:kmat}

In this section, we introduce k-matrices
for classical factorised scattering problems.
We provide several specific examples related to the algebra $\alg{sl}(2)$
on which our subsequent construction will rely.

\subsection{Factorised Scattering}

Consider a set of particles propagating along a one-dimensional space
whose flavour degrees of freedom are described by some vector space $\spc{V}$.
Each particle is specified by its momentum $p$ and a flavour $v\in\spc{V}$.
The pairwise scattering matrix $S(p_1,p_2):\spc{V}\otimes\spc{V}\to\spc{V}\otimes\spc{V}$
describes the scattering factors when two particles meet
and exchange their relative ordering along the one-dimensional space.
In case the one-dimensional space is a finite interval,
a reflection matrix $R(p):\spc{V}\to \spc{V}$
\footnote{One might assume the vector space $\spc{V}_{\text{out}}$ for reflected particles
to be different from the vector space $\spc{V}_{\text{in}}$ before reflection.
This case may be embedded into a reflection matrix
for the combined space $\spc{V}=\spc{V}_{\text{in}}\oplus\spc{V}_{\text{out}}$
and thus requires no special treatment.}
\footnote{If the boundary hosts dynamical degrees of freedom
which can interact with the bulk particles,
the reflection matrix extends to a two-particle scattering process
described by some operator
$R(p,p'):\spc{V}\otimes\spc{V}'\to\spc{V}\otimes\spc{V}'$
where $p'$ and $\spc{V}'$ describe the boundary degrees of freedom.
This work will entirely deal with the simpler setup
without boundary degrees of freedom.}
describes the scattering factors for reflection of a particle 
with momentum $p\to -p$ off a boundary.

For factorised scattering problems,
the two-particle scattering and one-particle reflection matrices suffice
to describe all interactions exhaustively.
They are often expressed as the quantum R-matrix
\unskip\footnote{This R-matrix $\Rmat(u_1,u_2)$
is not (directly) related to the above reflection matrix $R(p)$.} 
$\Rmat(u_1,u_2):\spc{V}\otimes\spc{V}\to\spc{V}\otimes\spc{V}$
and the quantum K-matrix $\Kmat(u):\spc{V}\to \spc{V}$,
respectively, of some integrable system.
The spatial momenta $p_{1,2}$ are given by means of some function $p(u)$
of rapidity variables $u_{1,2}\in\Complex$,
and a map $u\mapsto \bar u$ with $p(\bar u)=-p(u)$
describes the resulting rapidity of reflected particles.
Consistency of the factorised scattering problem implies 
certain relations for the R- and K-matrices:
They both satisfy an inversion relation
(potentially up to some scalar prefactors depending on the rapidities)
\[
\Rmat_{12}(u_1,u_2)^{-1}=\Rmat_{21}(u_2,u_1)
\&
\Kmat(u)^{-1}=\Kmat(\bar u)
\]
The R-matrix must obey the Yang--Baxter equation
\[
\Rmat_{12}(u_1,u_2)\Rmat_{13}(u_1,u_3)\Rmat_{23}(u_2,u_3)
=
\Rmat_{23}(u_2,u_3)\Rmat_{13}(u_1,u_3)\Rmat_{12}(u_1,u_2)
\],
and the K-matrix obeys the (left) reflection equation
\cite{Cherednik:1984vvp,Sklyanin:1988yz,Kulish:1992qb}
\[
\Kmat_1(u_1)
\Rmat_{21}(\bar u_2,u_1)
\Kmat_2(u_2)
\Rmat_{12}(u_1,u_2)
=
\Rmat_{21}(\bar u_2,\bar u_1) 
\Kmat_2(u_2)
\Rmat_{12}(\bar u_1,u_2)
\Kmat_1(u_1)
\]
%

\subsection{Classical Limit}

The construction and analysis of scattering and reflection matrices
simplify in the classical limit where many objects and relations linearise.

Suppose the R-matrix depends on a parameter $\hbar$
and there is a point $\hbar=0$ where it reduces to the identity.
Then, we can define a classical limit where many relations simplify.
The classical r-matrix $\rmat_{12}:\spc{V}\otimes\spc{V}\to\spc{V}\otimes\spc{V}$
is obtained as the leading-order deviation of the R-matrix from the identity
\[
\Rmat_{12}(u_1,u_2) = \id_{12} + \iunit\hbar \rmat_{12}(u_1,u_2) + \Order(\hbar^2)
\&
\rmat_{12}(u_1,u_2) = \lim_{\hbar\to 0} \frac{\Rmat_{12}(u_1,u_2)-\id_{12}}{\iunit\hbar}
\]
The inversion relation implies an anti-symmetric r-matrix
$\rmat_{21}(u_2,u_1)=-\rmat_{12}(u_1,u_2)$ 
(potentially up to an identity matrix times a function of $u_1,u_2$).
The Yang--Baxter relation reduces to its classical counterpart
\[
\comm!{\rmat_{12}(u_1,u_2)}{\rmat_{13}(u_1,u_3)}
+\comm!{\rmat_{12}(u_1,u_2)}{\rmat_{23}(u_2,u_3)}
+\comm!{\rmat_{13}(u_1,u_3)}{\rmat_{23}(u_2,u_3)}
=0
\]
Reflections typically remain non-trivial in the classical limit $\hbar\to0$,
consequently, the K-matrix $\Kmat(u)$ merely
reduces to its leading-order contribution $\kmat(u):\spc{V}\to\spc{V}$
\[
\Kmat(u)=k(u)+\Order(\hbar)
\&
\kmat(u) = \lim_{\hbar\to 0} \Kmat(u)
\]
Then, the reflection equation reduces to the classical counterpart
\<=+s@{eq:classreq}
\kmat_1(u_1)\rmat_{21}(\bar u_2,u_1)\kmat_2(u_2)
+\kmat_1(u_1)\kmat_2(u_2)\rmat_{12}(u_1,u_2)
\\;{}
-\rmat_{21}(\bar u_2,\bar u_1) \kmat_2(u_1)\kmat_1(u_1)
-\kmat_2(u_2)\rmat_{12}(\bar u_1,u_2)\kmat_1(u_1)
=0
\>
%

\subsection{\texorpdfstring{$\alg{sl}(2)$}{sl(2)} Examples}

A prime example for the above structures
is the R-matrix for the XXX model
which is based on the Lie algebra $\alg{sl}(2)$
with the representation $\rho$ acting on the space $\spc{V}=\Complex^2$.
It admits a family of K-matrices.
In our later constructions of more elaborate reflection matrices,
we will heavily rely on this example, and we shall thus discuss
the emerging structures in detail.

Let us first outline our notation related to the $\alg{sl}(2)$ algebra
which is spanned by three generators $\slgen^a$ with $a=0,\pm$.
The Lie brackets are normalised as
\[@{eq:Jcomm}
\comm{\slgen^0}{\slgen^\pm}=\pm\slgen^\pm
\&
\comm{\slgen^+}{\slgen^-}=-2\slgen^0
\]
The algebra has a quadratic invariant tensor
for which we choose the normalisation
\[@{eq:sl2cas}
\slgen_1\sldot\slgen_2 \defeq
-\slgen^0_1\slgen^0_2
+\half\slgen^+_1\slgen^-_2
+\half\slgen^-_1\slgen^+_2
\]
Here, the lower indices distinguish tensor sites on which the generators act.
We introduce the structure constants $\slstr^{ab}{}_c$
and the metric tensor $\slinv_{ab}$ along with its inverse $\slinv^{ab}$
\[
\slstr^{0\pm}{}_{\pm} \&= \mp \iunit 
\&
\slstr^{\pm\mp}{}_{0} \&= \pm 2\iunit
\&
\slinv_{00} = \slinv^{00} \&= -1
\&
\slinv_{\pm\mp} \&= \half
\&
\slinv^{\pm\mp} \&= 2
\]~
such that the algebra relations and quadratic invariant combine as
\[
\comm{\slgen^a}{\slgen^b}=\iunit\slstr^{ab}{}_c\slgen^c
\&
\slgen_1\sldot\slgen_2=c_{ab} \slgen^a_1 \slgen^b_2
\]
The representation $\rho$ on $\spc{V}$ is given by 
\[
\rho(\slgen^a) = \half c^{ab}\slpauli_b
\],
where the $\slpauli_a$ and $\sleps$ are a set of $2\times 2$ matrices
\<@{eq:pauli}
\slpauli_0 \&= \<* 1 & 0 \\ 0 & -1 \>
\&
\slpauli_+ \&= \<* 0 & \iunit \\ 0 & 0 \>
\&
\slpauli_- \&= \<* 0 & 0\\ \iunit & 0 \>
\&
\sleps \&= \<* 0 & +1 \\ -1 & 0 \>
\>

The R-matrix for the rational XXX model can be expressed as
\unskip\footnote{We include an unconventional prefactor function
in order to make the R-matrix manifestly satisfy the inversion relation
$\Rmat_{12}(u_1,u_2)^{-1}=\Rmat_{21}(u_2,u_1)$.
It makes no difference for most relations.}
\[
\Rmat_{12}(u_1,u_2)
=\sqrt{\frac{u_1-u_2-\iunit\hbar}{u_1-u_2+\iunit\hbar}}
\frac{1}{u_1-u_2-\iunit\hbar} \brk!{(u_1-u_2)\mathinner{\id}_{12}-\iunit\hbar\mathinner{\permop}_{12}}
\],
where `$\permop$' denotes the permutation operator on $\spc{V}\otimes\spc{V}$.
Alternatively, the permutation can be expressed in terms of the matrices $\slpauli_a$
via the identity $\slpauli_{1}\sldot\slpauli_{2}=\mathinner{\id}_{12}-2\mathinner{\permop}_{12}$.
A reflection flips the sign of the rapidity
\[
\bar u=2 \parfix-u
\]
For the sake of generality,
we have positioned the fixed point of the reflection at $u=\parfix$.
A common and convenient choice is $\parfix=0$.
The reflection matrix takes the general form \cite{Sklyanin:1988yz}
\unskip\footnote{The unconventional prefactor
enables a direct comparison for some specific relations
without having to resort to statements about equivalence.
Note that the square root introduces a sign ambiguity
which may have to be resolved according to the particular situation.}
\footnote{The choice of prefactor serves as a parity.}
\[@{eq:Krat}
\Kmat(u)=
\pm\sqrt{\frac{\parrefl-u+\parfix}{\parrefl+u-\parfix}}
\frac{1}{u-\parfix-\parrefl}\brk!{-(u-\parfix)\. \dirrefl^a\slpauli_a-\parrefl\mathinner{\id}}
\]
Here, $\dirrefl^a$ is a fixed unit 3-vector, 
$\dirrefl\sldot\dirrefl=-1$,
which describes the orientation of the reflection,
and $\parrefl\in\bar\Complex$ describes the singular points
$u=\parfix\pm \parrefl$
of the reflection matrix.
Altogether, the reflection is described by 4 degrees of freedom
in $\parfix,\parrefl$ and $\dirrefl^a$.

The classical rational r-matrix simplifies to
\[@{eq:rrat}
\rmat_{12}(u_1,u_2)
=
\frac{1}{u_1-u_2}\brk!{\half\mathinner{\id}_{12}-\mathinner{\permop}_{12}}
\&
\kmat(u)=\Kmat(u)
\]
The form of the classical k-matrix $\kmat$
is the same as for its quantum counterpart $\Kmat$
because the latter does not depend on $\hbar$.

\subsection{Quasi-Triangular Affine Lie Bialgebras}

The above framework for scattering matrices can be expressed
abstractly in terms of bialgebras. 
The quantum scattering problem uses quantum affine algebras,
which are particular quasi-triangular Hopf algebras
whose universal R-matrix embodies scattering R-matrices
and their relations in an abstract, representation-independent way,
see \cite{Chari:1994pz,Gomez:1996az}.
Likewise, reflection matrices can be formulated in terms of universal K-matrices
\cite{2015arXiv150706276B,Appel:2020zze,Appel:2022scp}.
Unfortunately, quantum affine algebras are rather difficult to work with,
and it is easier to analyse the corresponding problem in the classical limit
where the relevant algebra turns into a quasi-triangular Lie bialgebra.
Such an algebra can be properly quantised
\cite{1995q.alg.....6005E,1997q.alg.....1038E,1996q.alg....10030E}, 
which, at least formally,
leads to the recovery of the full quasi-triangular Hopf algebra structure.

In the classical limit, the quantum affine algebra
reduces to an affine Kac--Moody algebra $\alg{g}$
which is equipped with the structure of a quasi-triangular Lie bialgebra.
The universal R-matrix trivialises and the leading correction term
becomes the classical r-matrix $\urmat\in \alg{g}\otimes\alg{g}$.
The coboundary relation expresses
the cobracket in terms of a Lie bracket with the r-matrix
\[
\cobra_{12}(\gen{X}) = \liebr{\gen{X}_1+\gen{X}_2}{\urmat_{12}}
\]
Anti-symmetry of the cobracket follows from the requirement that
$\urmat_{12}+\urmat_{21}$ must be an invariant element of $\alg{g}\otimes\alg{g}$.
The classical Yang--Baxter relation then ensures
that the bialgebra is consistent
\[
\cybe{\urmat}{\urmat}_{123}
\defeq
\liebr{\urmat_{12}}{\urmat_{13}}
+\liebr{\urmat_{12}}{\urmat_{23}}
+\liebr{\urmat_{13}}{\urmat_{23}}
\wanteq 0
\]

The above examples are based on the affine $\alg{sl}(2)$ algebra
which is spanned by the terms $a(u)\slgen^a$
with some function $a(u)$
together with an affine derivation $\genD$ and a central charge $\genC$.
\unskip\footnote{We will skip relevant details
about the space of allowable functions
which is, depending on the purpose,
a suitable completion of the ring of Laurent polynomials $\Complex[u,u^{-1}]$ in the formal variable $u$.}
The algebraic relations of the affine algebra can be summarised as
\<
\liebr!{a(u)\slgen^0}{b(u)\slgen^0} \&= -\frac{1}{2\pi\iunit} \oint_\infty a(u)\diff{b(u)} \genC
\\
\liebr!{a(u)\slgen^0}{b(u)\slgen^\pm} \&= \pm a(u)b(u) \slgen^\pm
\\
\liebr!{a(u)\slgen^+}{b(u)\slgen^-} \&= -2a(u)b(u)\slgen^0+\frac{1}{\pi\iunit} \oint_\infty a(u)\diff{b(u)} \genC
\\
\liebr!{\genD}{b(u)\slgen^a} \&= b'(u)\slgen^a
\>
From here on, we shall disregard the central charge
as we are ultimately interested in evaluation representations
for which $\genC=0$.
The rational r-matrix takes the algebraic form
\[@{eq:urrat}
\urmat_{12} = \frac{2\slgen_1\sldot\slgen_2}{u_1-u_2}
\].
Note that the evaluation representation
$\rho(u)$ of $\alg{g}$ on $\spc{V}$ is given by
\[
\rho(u)\brk!{a(u)\slgen^a} = a(u) \rho(\slgen^a)
\]
The tensor product of two evaluation representations
readily leads to the above r-matrix \eqref{eq:rrat}
as a representation of \eqref{eq:urrat}.
\unskip\footnote{We refrain from representing the affine derivation
which shifts the rapidity variable $u$;
it would thus join the representations $\rho(u)$ for all $u$
into a single irreducible representation.
Moreover, $\rho(\genC)=0$ for evaluation representations.}

\subsection{Reflection Element}

Next, we would like to incorporate the classical reflection matrix
as an abstract element $\ukmat$,
which we shall call the k-matrix, 
into the framework of the quasi-triangular Lie bialgebra.
We shall need such an abstract k-matrix for the formulation and verification of the reduction method. 
A reflection is a large transformation,
hence it cannot be expressed as an element of some Lie algebra $\alg{g}$.
Instead, the k-matrix must be a group-like object.
We propose a self-consistent formulation
for an abstract k-matrix
with the following properties:
\unskip\footnote{The following discussion is compatible with
the picture that $\ukmat$ belongs to a Lie group $\grp{G}$
with corresponding Lie algebra $\alg{g}$.
In particular, $\ukmat$ is an involution element
which belongs to a component of $\grp{G}$
that is disconnected from the identity element.
A simple example of such a structure
is given by the Lie group $\grp{O}(n)$ and its Lie algebra $\alg{so}(n)$,
where the k-matrix describes a pure reflection.}

A k-matrix $\ukmat$ acts on elements of the Lie bialgebra as a $\Integer_2$-automorphism
\[@{eq:adprop}
\ad(\ukmat):\alg{g}\to\alg{g}
\&
\ad(\ukmat)\brk!{\liebr{a}{b}}
=
\liebr!{\ad(\ukmat)(a)}{\ad(\ukmat)(b)}
\&
\ad(\ukmat)^2=\id
\]
In particular, we assume that it acts on the affine algebra by reflecting
the rapidity variable with some involution map
\[
u \longleftrightarrow \bar u
\]
Compatibility between a reflection and scattering matrix 
is given by the classical reflection equation \eqref{eq:classreq}.
We can express the latter using the involution properties
and the adjoint action $\ad(A)B=ABA^{-1}$ at the level of representations
\<=+s
\rmat_{12}(u_1,u_2)
+\ad\brk!{\kmat_1(\bar u_1)\kmat_2(\bar u_1)}\rmat_{12}(\bar u_1,\bar u_2)
\\;{}
-\ad\brk!{\kmat_2(\bar u_2)}\rmat_{12}(u_1,\bar u_2)
-\ad\brk!{\kmat_1(\bar u_1)}\rmat_{12}(\bar u_1,u_2)
=0
\>
In this form we can re-express the reflection equation as an abstract relation 
between the k- and r-matrices:
\[@{eq:algrefeq}
\brk!{\id-\ad(\ukmat)}{}^{}_1\brk!{\id-\ad(\ukmat)}{}^{}_2\urmat_{12}=0
\]

The relations for concrete reflection matrices $\kmat(u)$
then follow as evaluation representations from abstract algebraic relations for $\ukmat$.
In the presence of a k-matrix, an evaluation representation $\rho(u)$
should be supplemented with the reflected evaluation representation $\rho(\bar u)$
into the direct sum representation $\rho|u|$ on $\spc{V}\oplus\spc{V}$.
\unskip\footnote{For the special case $\bar u=u$ one may find additional representations.}
The k-matrix is then represented by
\[@{eq:repk}
\rho|u|(\gen{X}) = \<* \rho(u)(\gen{X}) & 0 \\ 0 & \rho(\bar u)(\gen{X}) \> 
\\
\rho|u|(\ukmat) = \<* 0 & \kmat(\bar u) \\ \kmat(u) & 0 \> 
\]
This representation must be compatible with the adjoint action 
of $\ukmat$ on $\alg{g}$
\[@{eq:repk2}
\rho|u|\brk!{\ad(\ukmat)(\gen{X})} = \rho|u|(\ukmat)\.\rho|u|(\gen{X})\.\rho|u|(\ukmat)
=
\<* \kmat(\bar u)\.\rho(\bar u)(\gen{X})\.\kmat(u) & 0 \\ 0 & \kmat(u)\.\rho(u)(\gen{X})\.\kmat(\bar u) \> 
\]
Both non-trivial components of this equation are equivalent
to a single compatibility relation between the adjoint action
and the evaluation representations $\rho(u)$ and $\rho(\bar u)$:
\[@{eq:repk3}
\rho(\bar u)\brk!{\ad(\ukmat)(\gen{X})} 
= \kmat(u)\.\rho(u)(\gen{X})\.\kmat(\bar u)
\]
The adjoint action corresponding to the general rational $\alg{sl}(2)$ k-matrix \eqref{eq:rrat}
can be deduced from \eqref{eq:repk3}
by inverting the fundamental evaluation representation which is faithful.
It takes the form
\<@{eq:ukratact}
\ad(\ukmat_{\text{gen}})\brk!{a(u)\slgen^a} 
\&=
a(\bar u) \brk[s]*{\slgen^a +
\frac{
-2\iunit(u-\parfix)\parrefl\slstr^a{}_{bc}\dirrefl^b\slgen^c
-2(u-\parfix)^2(\slgen^a + \dirrefl^a \dirrefl\sldot\slgen)
}{(u-\parfix)^2-\parrefl^2}
}
\\
\ad(\ukmat_{\text{gen}})\brk{\genD}
\&= 
-\genD+\frac{2\parrefl\dirrefl\sldot\slgen}{(u-\parfix)^2-\parrefl^2}
\>
The adjoint action satisfies the axioms \eqref{eq:adprop},
and it is compatible with the r-matrix \eqref{eq:urrat}
via the reflection equation \eqref{eq:algrefeq}.

\subsection{Families of Reflection Elements}
\label{sec:deformations}

Part of the goal of this work is to obtain a complete enumeration 
of k-matrix structures suitable to enhance a given quasi-triangular Lie bialgebra.
A useful observation to this end
is that we can relate the general $\alg{sl}(2)$ k-matrix $\ukmat_{\text{gen}}$
to the basic k-matrix $\ukmat_0$
at $\parfix=0$, $\parrefl=\infty$ which is defined by
\unskip\footnote{Both choices for the sign of $\kmat_0$
are compatible with the given adjoint action $\ad(\ukmat_0)$.
Hence, the sign describes the parity of the representation $\rho|u|$
(which ought to have it as an additional label).}
\[
\ad(\ukmat_0)\brk!{a(u)\slgen^a} = a(-u) \slgen^a
\&
\ad(\ukmat_0)\brk{\genD} = -\genD
\&
\kmat_0(u)=\pm\id
\]
As $\ukmat_0$ incorporates an algebra automorphism,
we can obtain other automorphisms by combining them
with inner automorphisms generated by the exponential map
$\exp(\gen{X})$ with $\gen{X}$ some element of the affine Lie algebra.
In particular, the general $\alg{sl}(2)$ k-matrix defined in \eqref{eq:ukratact}
can be expressed conveniently as the shifted group element:
\<@{eq:ukrat}
\ukmat_{\text{gen}}
\&=
\ad\brk[s]!{
\exp\brk{-\parfix\genD}
\exp\brk!{
\log\brk*{1-u/\parrefl}\dirrefl\sldot\slgen
}}(\ukmat_0)
\\=
\ukmat_0
\exp\brk!{2\parfix\genD}
\exp\brk[s]*{
\log\brk*{\frac{\parrefl+u-\parfix}{\parrefl-u+\parfix}}
\dirrefl\sldot\slgen
}
\>
By construction, $\ukmat_{\text{gen}}$ describes a $\Integer_2$-automorphism of $\alg{g}$,
while the particular choice of algebra element turns out to restrict
to actual solutions of the reflection equation.
Its representation follows immediately
from the corresponding representation of the basic k-matrix:
\[
\kmat_{\text{gen}}(u) 
= \pm\exp\brk[s]*{\half\log\brk*{\frac{\parrefl+u-\parfix}{\parrefl-u+\parfix}} \dirrefl^a\slpauli_a}
\],
The equivalence of this expression to the earlier form in \eqref{eq:Krat}
follows by evaluation on the eigenvectors of $\dirrefl^a\slpauli_a$.

The form \eqref{eq:ukrat} has an interesting implication on the Lie groups
generated by $(\alg{g},\ukmat_{\text{gen}})$,
namely $\grp{G}_{\text{gen}}=\grp{G}_0\cup\ukmat_{\text{gen}}\grp{G}_0$
with $\grp{G}_0\defeq\exp\alg{g}$:
The exponentiated affine algebra elements describe inner automorphisms of $\alg{g}$
whereas $\ukmat_0$ serves as an outer automorphism for the affine algebra $\alg{g}$.
Therefore, $\ukmat_{\text{gen}}$ actually belongs to the Lie group
$\grp{G}\defeq\grp{G}_0\cup\ukmat_0\grp{G}_0$
generated by $\alg{g}$ together with the basic reflection $\ukmat_0$.
This shows that all Lie groups generated by $(\alg{g},\ukmat_{\text{gen}})$
for different values of the parameters $\parfix,\parrefl,\dirrefl^a$
are in fact equivalent groups, $\grp{G}_{\text{gen}}=\grp{G}$.
The odd part $\ukmat_0\grp{G}_0$ of the reflection group $\grp{G}$
contains all k-matrices $\ukmat_{\text{gen}}$ satisfying the reflection equation,
along with a much larger number of reflection elements which do not.

\medskip

More generally, we will express relevant continuous families of k-matrices
as the combination of a basic algebra automorphism generated
by some group-like object $\ukmat_0$
together with a family of $\ukmat_0$-odd algebra elements $\gen{X}\in\alg{g}$
\[
\ukmat
=\ad\brk[s]{\exp(\gen{X})}(\ukmat_0)
=\ukmat_0\exp(-2\gen{X})
\&
\ad(\ukmat_0)(\gen{X})=-\gen{X}
\]
This approach also allows us to study restrictions on allowable deformations $\ukmat$
of $\ukmat_0$ within its neighbourhood with the ansatz
\[
\ukmat=\ukmat_0\exp\brk!{\epsilon \gen{X}+\epsilon^2 \gen{X}'+\ldots}
\]
The algebra elements $\gen{X},\gen{X}'\in\alg{g}$ have to satisfy the
involution relations $\ad(\ukmat_0)\gen{X}=-\gen{X}$
and $\ad(\ukmat_0)\gen{X}'=-\gen{X}'$.
The allowable leading deformations $\gen{X}$ are restricted by the 
leading deformation of the reflection equation \eqref{eq:algrefeq},
\[@{eq:deformre}
\liebr!{\gen{X}_1}{\brk!{\id-\ad(\ukmat_0)}{}^{}_2\urmat_{12}}
+\liebr!{\gen{X}_2}{\brk!{\id-\ad(\ukmat_0)}{}^{}_1\urmat_{12}}
=0
\]
Unfortunately, this condition is not sufficient to guarantee that $\gen{X}$
actually generates a valid deformation:
For instance, the second-order deformation of the reflection equation reads:
\[
\liebr!{\gen{X}'_1}{\brk!{\id-\ad(\ukmat_0)}{}^{}_2\urmat_{12}}
+\liebr!{\gen{X}'_2}{\brk!{\id-\ad(\ukmat_0)}{}^{}_1\urmat_{12}}
+\half
\liebr!{\gen{X}_1}{\liebr{\gen{X}_2}{
\ad(\ukmat_0){}^{}_1\urmat_{12}+\ad(\ukmat_0){}^{}_2\urmat_{12}}}=0
\]
Here, the quadratic combinations from $\gen{X}$ must compensate the
linear contribution from $\gen{X}'$.
This equation may or may not have a solution for all $\gen{X}$
satisfying the leading-order deformation equation.
Only those $\gen{X}$ satisfying both equations for some $\gen{X}'$
(and potentially higher-order consistency equations)
constitute valid deformations.
However, in the special case $\ad(\ukmat_0){}_1\urmat_{12}+\ad(\ukmat_0){}_2\urmat_{12}=0$,
\unskip\footnote{For example, this equality holds for the $\alg{sl}(2)$ r-matrix,
hence, there are no additional constraints at second order.}
the deformation equation suffices.

In conclusion, the deformation reflection equation \eqref{eq:deformre}
serves as a necessary condition for allowable deformations $\gen{X}$
to some k-matrix $\ukmat_0$,
but it is clearly not a sufficient condition in all cases.
The dimension of the solution space of this linear equation
puts an upper bound to the dimensionality of the family of k-matrices.

\subsection{Diagonal and Null Reflections}
\label{sec:diagnull}

Let us consider two particular cases of interest from the family
of general k-matrices.
A standard choice for the parameters in \eqref{eq:ukrat}
is the orientation $\dirrefl^a=\delta^{a,0}$ leading to a diagonal k-matrix
(with respect to the canonical Cartan element $\slgen^0$):
\[@{eq:ukratdiag}
\ukmat_{\text{diag}}
\&=
\ukmat_0
\exp\brk!{2\parfix\genD}
\exp\brk[s]*{\log\brk*{\frac{\parrefl-u+\parfix}{\parrefl+u-\parfix}}\slgen^0}
\]
More concretely, its adjoint action takes the form
\<@{eq:ukratdiagact}
\ad(\ukmat_{\text{diag}})\brk!{a(u)\slgen^0} \&= a(\bar u)\slgen^0
\\
\ad(\ukmat_{\text{diag}})\brk!{a(u)\slgen^\pm} \&= 
a(\bar u)
\brk*{\frac{\parrefl+u-\parfix}{\parrefl-u+\parfix}}^{\pm 1}
\slgen^\pm
\\
\ad(\ukmat_{\text{diag}})\brk{\genD} \&= 
-\genD-\frac{2\parrefl\slgen^0}{(u-\parfix)^2-\parrefl^2}
\>
In this case, the k-matrix representation reduces to 
\[
\kmat_{\text{diag}}(u) 
= \pm\sqrt{\frac{\parrefl-u}{\parrefl+u}}
\frac{1}{u-\parrefl}
\brk!{-u\slpauli_0-\parrefl\id}
\],
which is compatible with the representation of the above adjoint action.
At the points $\parrefl=\infty$ and $\parrefl=0$,
the k-matrix becomes independent of the rapidity variable $u$.
Furthermore, at $\parrefl=\infty$ it is isotropic and
at $\parrefl=0$ the reflection is along a fixed direction.

\medskip

Another case of interest is the limit
where the orientation vector $\dirrefl^a$ aligns with a null vector $\nullrefl^a$
with $\nullrefl\sldot\nullrefl=0$.
We can achieve the null orientation
by choosing a reference unit vector $\dirrefl_0^a$
orthogonal to $\nullrefl^b$
with $\dirrefl_0\sldot\dirrefl_0=-1$ and $\dirrefl_0\sldot\nullrefl=0$.
Then $\dirrefl^a = \dirrefl_0^a + \alpha \nullrefl^a$ is a suitable
orientation vector for all $\alpha$ which approaches a null vector for large $\alpha$.
The limit $\alpha\to\infty$ causes divergences unless $\parrefl$
is tuned to simultaneously approach $\infty$.
We parametrise the orientation vector as a function of $\parrefl$ and
the null vector $\nullrefl^a$ as follows
\[@{eq:nulllimit}
\dirrefl^a=\dirrefl_0^a + \parrefl \nullrefl^a
\&
\parrefl\to\infty
\]
The limiting k-matrix \eqref{eq:ukrat} reads
\<@{eq:ukratnull}
\ukmat_{\text{null}}
\&=
\ukmat_0
\exp\brk!{2\parfix\genD}
\exp\brk!{2(u-\parfix)\nullrefl\sldot\slgen}
\>~
with the limiting adjoint action \eqref{eq:ukratact}
\<
\ad(\ukmat_{\text{null}})\brk!{a(u)\slgen^a} 
\&=
a(\bar u) \brk[s]*{\slgen^a 
+2\iunit(u-\parfix)\slstr^a{}_{bc}\nullrefl^b\slgen^c
+2(u-\parfix)^2\nullrefl^a \nullrefl\sldot\slgen
}
\\
\ad(\ukmat_{\text{null}})\brk{\genD}
\&= 
-\genD-2\nullrefl\sldot\slgen
\>
A standard choice is the null direction $\nullrefl^a=\delta^{a,-}\nullrefl$.
For it, the adjoint action of the k-matrix takes an upper-triangular form
\<@{eq:ukratnullact}
\ad(\ukmat_{\text{null}})\brk!{a(u)\slgen^+} \&= a(\bar u)\slgen^+
\\
\ad(\ukmat_{\text{null}})\brk!{a(u)\slgen^0} \&= a(\bar u)
\brk!{\slgen^0 + \nullrefl(u-\parfix) \slgen^+}
\\
\ad(\ukmat_{\text{null}})\brk!{a(u)\slgen^-} \&= a(\bar u)
\brk!{\slgen^- +2\nullrefl(u-\parfix) \slgen^0 + \nullrefl^2(u-\parfix)^2 \slgen^+}
\\
\ad(\ukmat_{\text{null}})\brk{\genD}
\&= 
-\genD-\nullrefl \slgen^+
\>,
and it has the representation
\[
\kmat_{\text{null}}(u) 
= \pm\brk!{\mathinner{\id}+\nullrefl(u-\parfix)\slpauli_-}
\]

\subsection{Null Twist}
\label{sec:nulltwist}

A generalisation relevant to our later study 
concerns twist-\hspace{0pt}deformations of the standard r-matrix,
see \cite{stolin1991}.
Let us briefly expose the effect of such a deformation on the corresponding k-matrices.
The null twist for the above standard r-matrix \eqref{eq:urrat} takes the form 
\[
\urmat^{\text{twist}}_{12} = -\nu\frac{\slgen_1\sldot\slgen_2}{u_1-u_2}
+
\iunit \dirtw^a\slstr_{abc} \slgen^b_1\slgen^c_2
\],
with the parameter $\dirtw^a$ an arbitrary null vector,
$\dirtw\sldot\dirtw=0$.
For instance, the direction $\dirtw^a=2\delta^{-,a}\dirtw$ yields the standard form 
$\iunit\dirtw^a\slstr_{abc}\slgen^b\otimes\slgen^c=\dirtw \slgen^0\wedge\slgen^+$.
This deformation satisfies the classical Yang--Baxter equation.
The twist appears in the reflection equation 
and the k-matrix solutions correspondingly picks up the following dependency on $\dirtw^a$:
\[@{eq:ukrattw}
\ukmat_{\text{tw}}
\&=
\ukmat_0
\exp\brk!{2\parfix\genD}
\exp\brk[s]*{
\log\brk*{\frac{\parrefl+u-\parfix+\nu^{-1}\dirtw\sldot\dirrefl (u-\parfix)^2}
               {\parrefl-u+\parfix+\nu^{-1}\dirtw\sldot\dirrefl (u-\parfix)^2}}
\dirrefl\sldot\slgen
}
\]
For a null reflection, we find the limit
\[@{eq:ukrattwnull}
\ukmat_{\text{twist},\text{null}}
\&=
\ukmat_0
\exp\brk!{2\parfix\genD}
\exp\brk[s]*{
\frac{2(u-\parfix)}{1+\nu^{-1}\dirtw\sldot\nullrefl (u-\parfix)^2}\nullrefl\sldot\slgen
}
\]
We note that for $\dirtw\sldot\nullrefl\neq0$, this expression has 
a further special case in the limit of a long null vector 
which only depends on the direction of $\nullrefl$
\[
\ukmat_{\text{twist},\text{null}}
\&=
\ukmat_0
\exp\brk!{2\parfix\genD}
\exp\brk[s]*{
\frac{2\nu}{\dirtw\sldot\nullrefl (u-\parfix)}\nullrefl\sldot\slgen
}
\]
%

\section{Contraction and Reduction}
\label{sec:contractreduct}

In this section, we construct the family of reflection matrices
that are compatible with the contraction and reduction procedure
introduced in \cite{Beisert:2022vnc}.
The procedure starts with the loop bialgebra of $\alg{so}(2,2)=\alg{sl}(2)\times\alg{sl}(2)$ 
which constitutes the isometries of 3D AdS spacetime.
This algebra is subsequently contracted to the loop algebra
of the 3D Poincaré algebra $\alg{iso}(2,1)=\alg{sl}(2)\ltimes\Real^{2,1}$.
Finally, the algebra is reduced to a subalgebra from which an ideal is removed
to yield a deformation of the loop algebra of $\alg{gl}(1)\times\Real$.
\[`{}@{eq:fig-bosonic}
\begin{mpostfig}[opt=align]
vpos[1]:=(0,0);
vpos[2]:=vpos[1]+(3.5cm,0);
vpos[3]:=vpos[2]+(4.5cm,0);
vpos[4]:=vpos[3]+(4cm,0);
paths[1]:=rbox(vpos[1],(1.4cm,0.7cm),5pt);
paths[2]:=rbox(vpos[2],(2.3cm,0.7cm),5pt);
paths[3]:=rbox(vpos[3],(2cm,0.7cm),5pt);
paths[4]:=rbox(vpos[4],(1.9cm,0.7cm),5pt);
paths[12]:=(point 2.5 of paths[1])--(point 6.5 of paths[2]);
paths[13]:=(point 2.5 of paths[2])--(point 6.5 of paths[3]);
paths[14]:=(point 2.5 of paths[3])--(point 6.5 of paths[4]);
for i:=1 upto 4: fillshape(paths[i],0.9white); endfor
for i:=12 upto 14: drawarrow subpath(0.1,0.9) of paths[i]; endfor
label(btex $\alg{sl}(2)$ etex, center paths[1]);
label(btex $\alg{sl}(2)\times\alg{sl}(2)$ etex, center paths[2]);
label(btex $\alg{sl}(2)\ltimes\Real^{2,1}$ etex, center paths[3]);
label(btex $\alg{gl}(1)\ltimes\Real$ etex, center paths[4]);
label.top(btex pairing etex, point 0.5 of paths[12]);
label.top(btex contraction etex, point 0.5 of paths[13]);
label.top(btex reduction etex, point 0.5 of paths[14]);
\end{mpostfig}
\]
The r-matrix, which provides the quasi-triangular Lie bialgebra structures,
passes through the contraction step without further ado.
The reduction step, however, requires a particular Drinfeld twist
which we will add from the beginning.

The extension of the procedure to the k-matrix is slightly more involved
and requires the distinction of several cases for the k-matrix parameters
which also depend on the r-matrix twist.
We will start with a simplified setup without an r-matrix twist.
Here, the k-matrix features two main cases
because the associated automorphism may either map the
$\alg{sl}(2)$ factors to themselves or permute them:
\[`{}@{eq:fig-untwisted}
\begin{mpostfig}[opt=align]
nums[1]:=2cm;
vpos[1]:=(0,0);
vpos[2]:=vpos[1]+(nums[1]+2.5cm,+0.0cm);
vpos[3]:=vpos[2]+(nums[1]+2.5cm,0);
vpos[4]:=vpos[1]+(nums[1]+2.5cm,-0.8cm);
vpos[5]:=vpos[4]+(nums[1]+2.5cm,0);
paths[1]:=rbox(vpos[1],(nums[1],0.5cm),5pt);
paths[2]:=rbox(vpos[2],(nums[1],0.5cm),5pt);
paths[3]:=rbox(vpos[3],(nums[1],0.5cm),5pt);
paths[4]:=rbox(vpos[4],(nums[1],0.5cm),5pt);
paths[5]:=rbox(vpos[5],(nums[1],0.5cm),5pt);
paths[12]:=(point 2.5 of paths[1])--(point 6.5 of paths[2]);
paths[13]:=(point 2.5 of paths[2])--(point 6.5 of paths[3]);
paths[14]:=(point 2.5 of paths[1])--(point 6.5 of paths[4]);
paths[14]:=(point 2.5 of paths[4])--(point 6.5 of paths[5]);
for i:=1 upto 5: fillshape(paths[i],0.9white); endfor
for i:=12 upto 14: drawarrow subpath(0.1,0.9) of paths[i]; endfor
label.top(btex $\alg{sl}(2)$ etex, point 4.5 of paths[1]);
label.top(btex $\alg{sl}(2)\times\alg{sl}(2)$ etex, point 4.5 of paths[2]);
label.top(btex $\alg{sl}(2)\ltimes\Real^{2,1}$ etex, point 4.5 of paths[3]);
label(btex $\ukmat_{\text{gen}}$ etex, center paths[1]);
label(btex $\ukmat_{\text{gen}}\otimes\ukmat_{\text{gen}}$ etex, center paths[2]);
label(btex $\ukmat_{\text{gen}}$ etex, center paths[3]);
label(btex $\ukmat_{\text{perm}}$ etex, center paths[4]);
label(btex $\ukmat_{\text{perm}}$ etex, center paths[5]);
label.top(btex pairing etex, point 0.5 of paths[12]);
label.top(btex contraction etex, point 0.5 of paths[13]);
\end{mpostfig}
\]
Here, the contraction step is straight-forward for both cases,
but reduction must fail in the absence of twists.
We find that introduction of the twist imposes certain restrictions to
the orientation of admissible k-matrices.
These split the no permutation case into two separate subcases
while the permutation case merely has fewer degrees of freedom.
The contraction step is again passed without obstacles,
and the reduction step fixes further parameters
without further splitting the families:
\[`{}@{eq:fig-twisted}
\begin{mpostfig}[opt=align]
nums[1]:=2cm;
vpos[1]:=(0,0);
vpos[2]:=vpos[1]+(nums[1]+2.5cm,+0.4cm);
vpos[3]:=vpos[2]+(nums[1]+2.5cm,0);
vpos[4]:=vpos[3]+(nums[1]+2.5cm,0);
vpos[5]:=vpos[1]+(nums[1]+2.5cm,-0.4cm);
vpos[6]:=vpos[5]+(nums[1]+2.5cm,0);
vpos[7]:=vpos[6]+(nums[1]+2.5cm,0);
vpos[8]:=vpos[1]+(0,-1.2cm);
vpos[9]:=vpos[8]+(nums[1]+2.5cm,0);
vpos[10]:=vpos[9]+(nums[1]+2.5cm,0);
vpos[11]:=vpos[10]+(nums[1]+2.5cm,0);
vpos[12]:=vpos[11]+(0,-0.8cm);
for i:=1 upto 12: paths[i]:=rbox(vpos[i],(nums[1],0.5cm),5pt); endfor
paths[22]:=(point 2.5 of paths[1])--(point 6.5 of paths[2]);
paths[23]:=(point 2.5 of paths[2])--(point 6.5 of paths[3]);
paths[24]:=(point 2.5 of paths[3])--(point 6.5 of paths[4]);
paths[25]:=(point 2.5 of paths[1])--(point 6.5 of paths[5]);
paths[26]:=(point 2.5 of paths[5])--(point 6.5 of paths[6]);
paths[27]:=(point 2.5 of paths[6])--(point 6.5 of paths[7]);
paths[28]:=(point 2.5 of paths[8])--(point 6.5 of paths[9]);
paths[29]:=(point 2.5 of paths[9])--(point 6.5 of paths[10]);
paths[30]:=(point 2.5 of paths[10])--(point 6.5 of paths[11]);
for i:=1 upto 12: fillshape(paths[i],0.9white); endfor
for i:=22 upto 30: drawarrow subpath(0.1,0.9) of paths[i]; endfor
label.top(btex $\alg{sl}(2)\times\alg{sl}(2)$ etex, point 4.5 of paths[1]);
label.top(btex $\alg{sl}(2)\times\alg{sl}(2)$ etex, point 4.5 of paths[2]);
label.top(btex $\alg{sl}(2)\ltimes\Real^{2,1}$ etex, point 4.5 of paths[3]);
label.top(btex $\alg{gl}(1)\ltimes\Real$ etex, point 4.5 of paths[4]);
label(btex $\ukmat_{\text{gen}}\otimes\ukmat_{\text{gen}}$ etex, center paths[1]);
label(btex $\ukmat_{\text{diag}}\otimes\ukmat_{\text{tw}}$ etex, center paths[2]);
label(btex $\ukmat_{\text{diag}}$ etex, center paths[3]);
label(btex $\ukmat_{\text{diag}}$ etex, center paths[4]);
label(btex $\ukmat_{\text{gen}}\otimes\ukmat_{\text{null}}$ etex, center paths[5]);
label(btex $\ukmat_{\text{null}}$ etex, center paths[6]);
label(btex $\ukmat_{\text{null}}$ etex, center paths[7]);
label(btex $\ukmat_{\text{perm}}$ etex, center paths[8]);
label(btex $\ukmat_{\text{perm}}$ etex, center paths[9]);
label(btex $\ukmat_{\text{perm}}$ etex, center paths[10]);
label(btex $\ukmat_{\text{perm}}$ etex, center paths[11]);
label(btex $\ukmat_{\text{perm}^*}$ etex, center paths[12]);
label.top(btex twist etex, point 0.5 of paths[28]);
label.top(btex contraction etex, point 0.5 of paths[29]);
label.top(btex reduction etex, point 0.5 of paths[30]);
\end{mpostfig}
\]
We point out that the twist may equally be implemented after contraction,
and that we find one further k-matrix family for $\alg{gl}(1)\ltimes\Real$
which apparently has no preimage from reduction.

The main goal of this section is to outline the contraction and reduction 
procedure for k-matrices. 
We shall discuss in detail how each step constrains the parameter space
which eventually leads to three disconnected families of k-matrices.

As such, this section focuses on the kinematics of the reflection,
for which a simplified bosonic setup suffices.
The procedure can be enriched with supersymmetry
which extends the k-matrix to include flavour
and which provides the exceptional structure for the AdS/CFT worldsheet scattering problem.
This will be the subject of \secref{sec:supersymmetry}.

\subsection{Pairing}
\label{sec:pairing}

The construction starts with two copies of the affine $\alg{sl}(2)$ algebra
generated by $\genM^{\labA,a},\genM^{\labB,a}$
and a classical r-matrix 
with a null twist which also mixes the two $\alg{sl}(2)$ algebras
\[@{eq:rpair}
\urmat_{12} =
-\nu^\labA\frac{\genM^\labA_1\sldot\genM^\labA_2}{u_1-u_2}
-\nu^\labB\frac{\genM^\labB_1\sldot\genM^\labB_2}{u_1-u_2}
+\brk[s]!{\brk{\dirtw^{\labA\labB}\genM^{\labA,0}+\dirtw^\labB\genM^{\labB,0}}\wedge \genM^{\labB,+}}{}^{}_{12}
\]
%

\paragraph{No Permutation Case.}

To familiarise ourselves with the setup,
let us first remove the mixed twist, $\dirtw^{\labA\labB}=0$,
but keep the null twist for $\dirtw^\labB\genM^{\labB,0}\wedge \genM^{\labB,+}$
corresponding to the twist parameter $\dirtw^{\labB,a}=2\delta^{a,-}\dirtw^\labB$.
Evidently, an applicable reflection matrix is given by
the product of k-matrices for the two $\alg{sl}(2)$ subalgebras
in \eqref{eq:ukrat} and \eqref{eq:ukrattw}
\[@{eq:ukrattwistfactored}
\ukmat_{\text{gen}} = \ukmat^\labA_{\text{gen}} \ukmat^\labB_{\text{tw}}
\]
Each of the k-matrices contributes 4 degrees of freedom,
and there is no coupling between the two sets of parameters,
hence the resulting k-matrix has 8 degrees of freedom.

These are exhaustive in the sense that all deformations
in the vicinity of a generic point belong to the eight-parameter family:
Deformations correspond to continuous automorphisms. 
The inner automorphisms belong to either of the affine $\alg{sl}(2)$ algebras.
Consequently, they deform only the corresponding factor $\ukmat^\labA_{\text{gen}}$, $\ukmat^\labB_{\text{tw}}$
for which we know that 4 degrees of freedom exhaust the solutions of the reflection equation.
The same applies to the continuous outer automorphisms for each affine $\alg{sl}(2)$ algebra
which extend the affine derivation $\genD$ to a Witt/Virasoro algebra generated by $a(u)\genD$.
The outer automorphism exchanging the two affine $\alg{sl}(2)$ algebras
is discrete, not suitable for generating deformations,
and will be discussed later.

\paragraph{Diagonal and Null Cases.}

Next, we should incorporate the mixed twist term $\genM^{\labA,0}\wedge \genM^{\labB,+}$.
In the reflection equation, it will exclusively generate
tensor product terms which are mixed between the two algebra factors.
There is no possibility for cancellation with other terms,
therefore, the mixed twist term must cancel on its own
in the reflection equation \eqref{eq:algrefeq},
and the solution must be within the class
$\ukmat^{\labA}_{\text{gen}} \ukmat^\labB_{\text{tw}}$.
Full cancellation requires that one of its two legs
$\genM^{\labA,0}$ or $\genM^{\labB,+}$
must be invariant under the reflection $\ad(\ukmat)$.
Hence, there are two corresponding families of solutions:
On the one hand, $\genM^{\labA,0}$ commutes only with
deformations proportional to $\genM^{\labA,0}$,
and hence the diagonal choice $\dirrefl^{\labA,a}\sim\delta^{a,0}$ 
with the k-matrix $\ukmat^\labA_{\text{diag}}$ in \eqref{eq:ukratdiag}
solves the reflection equation
\[@{eq:ukratcontdiag}
\ukmat_{\text{diag}} = \ukmat^\labA_{\text{diag}} \ukmat^\labB_{\text{tw}}
\]
On the other hand, $\genM^{\labB,+}$ commutes only with 
deformations proportional to $\genM^{\labB,+}$
leading to the null k-matrix factor $\ukmat^\labB_{\text{twist},\text{null}}$
with $\nullrefl^{\labB,a}\sim \delta^{a,-}$,
which reduces to $\ukmat^\labB_{\text{null}}$ in \eqref{eq:ukratnull} due to $\dirtw^\labB\sldot\nullrefl^\labB=0$,
\[@{eq:ukratcontnull}
\ukmat_{\text{null}} = \ukmat^\labA_{\text{gen}} \ukmat^\labB_{\text{null}}
\]
Each of these two restrictions removes two degrees of freedom from the
general solution. 
Altogether, we thus find two six-parameter families of k-matrices
of the above kinds.

\paragraph{Permutation Case.}

As previously mentioned,
an additional outer automorphism for our algebra
is given by the exchange of the two $\alg{sl}(2)$ factors.
Let us define the plain permutation as the element $\ukmat_{\text{perm},0}$
which acts as
\[@{eq:ukpermbasic}
\ad(\ukmat_{\text{perm},0})\brk!{a(u)\genM^{\labAB,a}} \&= a(\sigma^{\pm1} u)\genM^{\labBA,a}
\&
\ad(\ukmat_{\text{perm},0})\brk!{\genD^\labAB} \&= \sigma^{\mp1}\genD^\labBA
\]
Here, we have included an adjustable rescaling $\sigma$ of the rapidity variable 
between the two $\alg{sl}(2)$ factors.
\unskip\footnote{In principle, we might incorporate an arbitrary 
invertible transformation between the rapidity variables
for the two $\alg{sl}(2)$ factors.
Alternatively, these can be generated by automorphisms
based on $a(u)(\genD^\labA-\genD^\labB)$.}
The above k-matrix solves the reflection equation for the untwisted r-matrix
provided that the factor $\sigma$ compensates for the different
prefactors of the composite $\alg{sl}(2)\times\alg{sl}(2)$ r-matrix
\[@{eq:permrescale}
\sigma=-\frac{\nu^\labA}{\nu^\labB}
\]
We find that the following family of deformations preserves the reflection equation
\[@{eq:ukperm}
\ukmat_{\text{perm}}=
\ukmat_{\text{perm},0}
\exp\brk!{-\parfix\genD^\labA+\sigma^{-1}\parfix\genD^\labB
+\parrot_a \genM^{\labA,a}-\parrot_a \genM^{\labB,a}}
\]
It has 4 degrees of freedom given by the parameter $\parfix$
describing a relative shift of the rapidity variable
and the vector $\parrot_a$ describing a rotation between the two subalgebras.
The existence of additional deformations can be excluded by the deformation reflection equation \eqref{eq:deformre}.

We note that the deformation is independent of the rapidity $u$.
This restriction is explained as follows:
In the permutation case with an overall r-matrix
that is the sum of two r-matrices for the subalgebras,
the reflection equation \eqref{eq:algrefeq} splits into its even and odd parts
\[
\urmat_{12}+\ad(\ukmat)_1\ad(\ukmat)_2\urmat_{12}=
\ad(\ukmat)_1\urmat_{12}+\ad(\ukmat)_2\urmat_{12}=0
\], 
since they belong to different subalgebra spaces.
This implies that both constituent r-matrices
are related by an automorphism (of the affine algebra).
This automorphism together with its inverse forms the permutation k-matrix.
The untwisted rational r-matrix has a shift and a rotation symmetry
which appear as the 4 degrees of freedom for the k-matrix.

Accommodating the twist for the permutation case
turns out to require the restriction $\dirtw^{\labA\labB}=\dirtw^{\labB}$
for the twist parameters and the constraint $\parrot_-=0$ for the k-matrix. 
One can then show that all components of the twist deformation
cancel among themselves in the reflection equation.
Hence, the twisted r-matrix admits
a residual three-parameter family of permutation k-matrices $\ukmat_{\text{perm}}$.

\paragraph{Summary.} 

Let us summarise these results.
For the r-matrix without the mixed twist, we obtain two distinct families of k-matrices:
one which maps each $\alg{sl}(2)$ factor to itself (with 8 degrees of freedom)
and one which permutes both $\alg{sl}(2)$ factors (with 4 degrees of freedom).
The inclusion of the mixed twist then splits the first family into two, 
namely $\ukmat_{\text{diag}}$ and $\ukmat_{\text{null}}$
and reduces the number of degrees of freedom by two for each. 
Conversely, the k-matrix $\ukmat_{\text{perm}}$ of the permutation case
loses one degree of freedom due to the mixed twist.

\subsection{Contraction}
\label{sec:contraction}

The 3D Poincaré algebra $\alg{sl}(2)\ltimes\Complex^3$
is obtained as a contraction of the 3D AdS algebra $\alg{sl}(2)\times\alg{sl}(2)$.
The same applies to the corresponding affine extensions, 
albeit the affine 3D Poincaré algebra 
has two affine derivations and two central charges
\cite{Beisert:2024rke}.
The contraction readily extends to the r-matrix,
and the amount of independent parameters is maintained.
We will show that the k-matrix also has a proper contraction
and the number of allowable deformation parameters is preserved.

\paragraph{Bialgebra Contraction.}

The contraction aligns both $\alg{sl}(2)$ algebra factors
into a diagonal $\alg{sl}(2)$ algebra serving as the Lorentz rotations $\genL^a$
and differences becoming the momentum generators $\genP^a$
\footnote{In principle, one might rescale the generators $\genM_\labB$
depending on the loop level before matching them with the corresponding $\genM_\labA$.
This is equivalent to rescaling the prefactor $\nu^\labB$ in the r-matrix.
We can thus constrain the form of the contraction without loss of generality.}
\[@{eq:contraction}
\genM_\labA^a=\frac{1}{\epsilon\mref}\genP^a
\&
\genM_\labB^a=\genL^a-\frac{1}{\epsilon\mref}\genP^a
\]
The resulting algebra relations 
$\liebr{\genL}{\genL}\sim\genL$ and $\liebr{\genL}{\genP}\sim\genP$
in the contraction limit $\epsilon\to 0$
are those of a Poincaré algebra.
The affine derivation is transformed as
\[
\genDL=\genD_\labA+\genD_\labB
\&
\genDP=\epsilon\mref\genD_\labA
\]
Effectively, this implies a slightly different transformation 
for the two rapidity variables
\[
u^\labA=u
\&
u^\labB=u-\epsilon\mref v
\]
where $\genDL$ and $\genDP$
serve as derivatives on $u$ and $v$, respectively.
We will use this transformation to determine the proper contraction
for special points in the rapidity plane.

The r-matrix coefficients have to be oppositely aligned
to cancel a potential divergence in the contraction
\<
\nu^\labA \&= \epsilon \mref \nu^\labL
\&
\nu^\labB \&= -\epsilon \mref \nu^\labL + \epsilon^2\mref^2 \nu^\labP
\\
\dirtw^{\labA\labB} \&= -\epsilon \mref \dirtw^\labL 
\&
\dirtw^\labB \&= -\epsilon \mref \dirtw^\labL + \epsilon^2\mref^2 \dirtw^\labP
\>
The contraction limit of the r-matrix is
\<@{eq:urmatpoincare}
\urmat_{12} \&=
-\nu^\labL\frac{\genL_1\sldot\genP_2+\genL_2\sldot\genP_1}{u_1-u_2}
-\nu^\labP\frac{\genP_1\sldot\genP_2}{u_1-u_2}
+\dirtw^\labL\brk!{\genL^0\wedge \genP^+}{}^{}_{12}
+\dirtw^\labP\brk!{\genP^0\wedge \genP^+}{}^{}_{12}
\>
%

\paragraph{No Permutation Case.}

We will now consider the contraction for the various families of k-matrices
staring with those which do not permute the two $\alg{sl}(2)$ factors.
The contraction of the basic reflection $\ukmat_0$,
which merely flips the rapidity $u\to -u$, 
remains a simple reflection
\<
\ad(\ukmat_0)\brk!{a(u)\genL^a} \&= a(-u) \genL^a
\&
\ad(\ukmat_0)\brk{\genDL} \&= -\genDL
\\
\ad(\ukmat_0)\brk!{a(u)\genP^a} \&= a(-u) \genP^a
\&
\ad(\ukmat_0)\brk{\genDP} \&= -\genDP
\>
The k-matrix families $\ukmat_{\text{gen}}$, $\ukmat_{\text{diag}}$, $\ukmat_{\text{null}}$
contract to deformations of this basic reflection,
and the deformation element of the contracted k-matrix
is the contraction of the deformation element of the uncontracted k-matrix.

Let us first consider the absence of twists, $\dirtw^\labL=\dirtw^\labP=0$
because it demonstrates most clearly how the contraction acts on the k-matrix parameters.
In order to perform the contraction limit of two k-matrices of the form \eqref{eq:ukrat}
but with different parameters,
we have to align their parameters up to a small difference
\<
\parfix^\labA \&= \parfix
\&
\parfix^\labB \&= \parfix - \epsilon \mref \parfixP+\ldots
\\
\parrefl^\labA \&= \parreflL
\&
\parrefl^\labB \&= \parreflL - \epsilon \mref \parreflP+\ldots
\\
\dirrefl^{\labA,a} \&= \dirreflL^a
\&
\dirrefl^{\labB,a} \&= \dirreflL^a -\epsilon\mref \dirreflP^a+\ldots
\eqnsep
\dirreflL\sldot\dirreflP=0
\>
The resulting adjoint action of the k-matrix 
on $\genL^a,\genDL$ and $\genP^a,\genDP$
is analogous to the action on $\slgen^a,\genD$ 
given in \eqref{eq:ukratact}.
Note that the action on $\genL^a,\genDL$
has additional contributions proportional to $\genP^a$
due to the deformations proportional to $\genP^a,\genDP$.
We refrain from spelling them out because
they are lengthy and will not be relevant for us.

\paragraph{Diagonal Case.}

After having seen the effect of the contraction on the k-matrix,
we activate the twist \eqref{eq:rpair} in the r-matrix.
As we have seen, this yields two separate cases before contraction
which lead to two separate cases after contraction. We start with the diagonal case.

For the diagonal case, we choose $\dirreflL^a = \delta^{a,0}$
and a vector $\dirreflP^a$ with $\dirreflP^0=0$.
The $\epsilon\to 0$ limit of the deformation \eqref{eq:ukrattw}
then yields the contracted k-matrix
\<@{eq:ukratdiagcont}
\ukmat_{\text{diag}}
\&=
\ukmat_0
\exp\brk!{2\parfix\genDL+2\parfixP\genDP}
\exp\brk[s]*{
\log\brk*{\frac{\parrefl-u+\parfix}{\parrefl+u-\parfix}}\genL^0
}
\\;\quad\cdot
\exp\brk[s]^4[
-\frac{u-\parfix}{\parrefl+u-\parfix}\dirreflP^+\genP^-
-\frac{u-\parfix}{\parrefl-u+\parfix}\dirreflP^-\genP^+
\\;\qquad\qquad+
2\frac{\parfixP\parrefl+\parreflP(u-\parfix)+\nu^{-1}_\labL\dirtw^\labL\dirreflP^+ (u-\parfix)^3}
{(u-\parfix)^2-\parrefl^2}
\genP^0
\brk]
\>
Here, we have used the following exponential identity to disentangle
the contributions from the generators $\genL$ and $\genP$:
\[
\comm{X}{Y}=Y
\eqnjoin*{\implies}
\exp\brk!{a(X+Y)}=\exp(aX)\exp\brk!{(1-\econst^{-a})Y}
\]
Altogether, the k-matrix is determined by the 4 parameters
$\parrefl,\parreflP$ and $\dirreflP^\pm$
next to the 2 fixed-point parameters $\parfix,\parfixP$
resulting in 6 deformation parameters in total.
The adjoint action of the k-matrix 
on $\genL^a,\genDL$ and $\genP^a,\genDP$
is analogous to the action on $\slgen^a,\genD$ 
given in \eqref{eq:ukratdiagact}
with additional terms for $\genL^a,\genDL$ proportional to $\genP^a,\genDP$.
%
%

\paragraph{Null Case.}

For the null k-matrix $\ukmat_{\text{null}}$
we have to align the general untwisted k-matrix $\ukmat^\labA_{\text{gen}}$ in \eqref{eq:ukrat}
with the null k-matrix $\ukmat^\labB_{\text{null}}$ in \eqref{eq:ukratnull}.
Therefore the parameters for $\ukmat^\labA_{\text{gen}}$
have to be tuned to align with \eqref{eq:nulllimit}.
We find that the resulting k-matrix has 6 parameters as well:
a number $\nullrefl$ and an unconstrained vector $\nullrefl_\labP^a$
next to the fixed-point parameters $\parfix,\parfixP$.
The contracted k-matrix takes the form
\<@{eq:ukratnullcont}
\ukmat_{\text{null}}
=
\ukmat_0
\exp\brk!{2\parfix\genDL+2\parfixP\genDP}
\exp\brk[s]*{(u-\parfix)\brk!{\nullrefl\genL^++\nullrefl_\labP\sldot\genP}
-\parfixP\nullrefl\genP^+
}
\>
Again, the adjoint action of the k-matrix 
on $\genL^a,\genDL$ and $\genP^a,\genDP$
is analogous to the action on $\slgen^a,\genD$ 
given in \eqref{eq:ukratnullact}
up to contributions due to the deformations proportional to $\genP^a,\genDP$
in \eqref{eq:ukratnullcont}.

\paragraph{Permutation Case.}

We finally consider the case where the k-matrix permutes
the two $\alg{sl}(2)$ factors. As before, we start in the absence of a twist.

In the permutation case, a proper contraction limit requires
that the adjoint action of the k-matrix maps $\genM^{\labA,a}+\genM^{\labB,a}$
to a generator of the approximate form $\genM^{\labA,b}+\genM^{\labB,b}$.
However, the individual generators $\genM^{\labA,a}$ and $\genM^{\labB,a}$
transform with mutually inverse $\grp{SL}(2)$ elements due to \eqref{eq:ukperm},
so this transformation must approximate an involution.
Later on, the reduction step will only work if the transformation
is approximately the identity; we will this focus on this case.
We also have to account for the
transformation of rapidity variables \eqref{eq:permrescale} by the factor
\[
\sigma^{\pm 1}=1 \pm \epsilon\mref \frac{\nu^\labP}{\nu^\labL}+\ldots
\]
The contraction limit leads to the following adjoint action
of the basic permutation k-matrix $\ukmat_{\text{perm},0}$ in \eqref{eq:ukpermbasic}
\<
\ad(\ukmat_{\text{perm},0})\brk!{a(u)\genL^a} \&= a(u)\genL^a -2u a'(u) \frac{\nu^\labP}{\nu^\labL} \genP^a
\&
\ad(\ukmat_{\text{perm},0})\brk!{\genDL} \&= \genDL + 2\frac{\nu^\labP}{\nu^\labL}\genDP
\\
\ad(\ukmat_{\text{perm},0})\brk!{a(u)\genP^a} \&= -a(u) \genP^a
\&
\ad(\ukmat_{\text{perm},0})\brk{\genDP} \&= -\genDP
\>
The deformation \eqref{eq:ukperm} must be chosen to be small
for it to be compatible with the contraction
\[
\parfix=-\epsilon\mref \parfixP
\&
\parrot_a=-\epsilon\mref \lambda_a
\]
The resulting contracted k-matrix thus reads
\[@{eq:ukpermLP}
\ukmat_{\text{perm}}= \ukmat_{\text{perm},0} \exp\brk!{2\parfixP\genDP-2\lambda_a \genP^a}
\]

We now turn on the twist.
Before the contraction, compatibility with the twist imposed the constraint $\parrot_-=0$.
The contraction limit of this condition is $\lambda_-=0$
which is exactly what is required by compatibility with the twist after contraction.
In this case, we write out the additional deformation terms
in the adjoint action on $\genL^a$:
\<
\ad(\ukmat_{\text{perm}})\brk!{a(u)\genL^+} \&= 
a(u) \brk!{\genL^+ - 2\lambda_0 \genP^+}
-2 a'(u) \brk[s]*{u\frac{\nu^\labP}{\nu^\labL}+\parfixP} \genP^+
\\
\ad(\ukmat_{\text{perm}})\brk!{a(u)\genL^0} \&= 
a(u) \brk!{\genL^0 + 2\lambda_+\genP^+}
-2 a'(u) \brk[s]*{u\frac{\nu^\labP}{\nu^\labL}+\parfixP} \genP^0
\\
\ad(\ukmat_{\text{perm}})\brk!{a(u)\genL^-} \&= 
a(u) \brk!{\genL^- + 2\lambda_0 \genP^- + 4\lambda_+\genP^0}
-2 a'(u) \brk[s]*{u\frac{\nu^\labP}{\nu^\labL}+\parfixP} \genP^-
\>

\paragraph{Summary.}

Altogether, the contraction acts consistently
on the all three families of k-matrices,
and we obtain the same number of degrees of freedom.
More concretely,
we find two 6-parameter families mapping the $\alg{sl}(2)$ factors to themselves 
and one 3-parameter family permuting the $\alg{sl}(2)$ factors.
All k-matrices for the contracted algebra
can be obtained by contraction of those in the uncontracted algebra.

\subsection{Reduction}
\label{sec:reduction}

Finally, we consider the reduction of the 
affine Poincaré algebra to the affine algebra
based on one Lorentz rotation $\genL$ and one translation $\genP$.
The relevant criterion for the k-matrix is that
it remains compatible with the resulting bialgebra structures.

\paragraph{Bialgebra Reduction.}

The reduction procedure consists of two steps.
The first step is to restrict to a one-dimensional subalgebra
of the $\alg{sl}(2)$ Lorentz algebra.
Importantly the basic direction $\genLt$
is made to depend non-trivially on the rapidity variable
\[
\genLt\defeq 
\frac{u}{\beta}\genL^0
-\half\econst^{-\iunit\alpha} \genL^+
-\half\econst^{+\iunit\alpha} \genL^-
\]
Correspondingly, the affine derivation has to be adjusted such that
its algebra with $\genLt$ closes
\[
\genDLt = \genDL + \frac{\beta}{u^2-\beta^2}\brk!{\half\econst^{-\iunit\alpha}\genL^+ - \half\econst^{+\iunit\alpha}\genL^-}
\]
The restriction to the subalgebra induces two ideals among the momentum generators
\[
\gen{I}^{\pm}=\gen{P}^\pm - \econst^{\pm\iunit\alpha}\frac{\beta}{u} \genP^0
\]
We can factor out this ideal by setting $\gen{I}^\pm=0$.
Effectively this identifies the momentum vector $\genP^a$
in terms of a single momentum generator $\genPt$:
\[
\genP^0 = \frac{u}{\beta}\gen{P}
\&
\genP^\pm = \econst^{\pm\iunit\alpha} \gen{P}
\]
Upon setting $\dirtw^\labL=\econst^{-\iunit\alpha}\nu^\labL/\beta$
the twisted Poincaré r-matrix in \eqref{eq:urmatpoincare} reduces to 
\<
\urmat_{12} \&=
\nu^\labL\frac{\genL_1\genP_2+\genL_2\genP_1}{u_1-u_2}
+\frac{\nu^\labP}{\beta^2}\frac{u_1u_2-\beta^2}{u_1-u_2}\genP_1\genP_2
+\frac{\dirtw^\labP}{\beta}(u_1-u_2)\genP_1\genP_2
\>

Compatibility of a k-matrix with the reduction depends on two issues.
First, the adjoint action $\ad(\ukmat)$
has to map the subalgebra of the first step to itself.
Second, the adjoint action has to map the ideal to itself.
This ensures that the k-matrix acts properly within
the reduced algebra and remains a solution of the algebraic reflection equation.

\paragraph{Diagonal Case.}

Let us first consider the family of diagonal k-matrices $\ukmat_{\text{diag}}$.
The general adjoint action \eqref{eq:ukratdiagact} acts rather differently
on $\genL^0$ and $\genL^\pm$ producing different additional poles in the rapidity variable.
Consequently, the adjoint action on $\genLt$ cannot be proportional to $\genLt$.
This mismatch can be avoided by shifting these poles to zero or to infinity.
The latter case with $\parrefl=\infty$ leads to an isotropic action $\genL^a$, 
but it flips the sign for $u$ such that the components of $\genLt$
transform differently.
The former case with $\parrefl=\parfix=0$, 
however, is perfectly suitable because
\[
\ukmat_{\text{diag}}(u\genL^0) = (-u)\genL^0+\ldots
\&
\ukmat_{\text{diag}}(\genL^\pm) = -\genL^\pm+\ldots
\]
Similarly, all terms in the adjusted affine derivation $\genDLt$
and in the momentum ideals $\gen{I}^\pm$ flip sign.
Hence, we have to restrict to $\parrefl=\parfix=0$.
Importantly, the deformations that turn $\genL^a$ and $\genDL$
to some combination of the $\genP^b$
do not play a role towards compatibility
because all the $\genP^b$ belong to the subalgebra.
Altogether, we find the following adjoint action on the reduced generators
including all deformation terms
\<@{eq:redaddiag}
\ad(\ukmat_{\text{diag}})\brk!{a(u)\genLt} \&= 
-a(-u)\genLt
-2\parfixP \frac{u^2-\beta^2}{\beta^2} \brk[s]*{ a'(-u) - \frac{ua(-u)}{u^2-\beta^2}}\genPt
\\
\ad(\ukmat_{\text{diag}})\brk!{a(u)\genPt} \&= -a(-u)\genPt
\\
\ad(\ukmat_{\text{diag}})\brk!{\genDLt} \&= -\genDLt 
- \brk[s]*{\lambda'(u)-\frac{u\lambda(u)}{u^2-\beta^2}}\genP
\\
\ad(\ukmat_{\text{diag}})\brk!{\genDP} \&= -\genDP 
\>
The effect of the deformations $\parrefl$ and $\lambda^\pm$
is condensed in the function $\lambda(u)$
which has terms proportional to $1$ and $u^2$.
Altogether, the reduced k-matrix can be expressed as
\[
\ukmat_{\text{diag}}
=
\ukmat_{\text{diag},0}
\exp\brk!{2\parfixP\genDP+\lambda(u)\genPt}
\]
More generally, any even function $\lambda(u)$ 
is a permissible deformation for the reduced algebra.
We thus find that the k-matrix $\ukmat_{\text{diag}}$
for the reduced algebra has more allowable deformations
than the reduction of the k-matrix.
However, their effect is limited to the adjoint action of the affine derivation $\genDLt$.

\paragraph{Null Case.}

For the case of the null k-matrix $\ukmat_{\text{null}}$,
we have to restrict the coefficients of the adjoint action in \eqref{eq:ukratnullact} 
in order to preserve the form of $\genLt$.
This is achieved by fixing $\nullrefl=-2\econst^{-\iunit\alpha}/\beta$ and $\parfix=0$.
The full adjoint action of the reduced k-matrix reads
\<@{eq:redadnull}
\ad(\ukmat_{\text{null}})\brk!{a(u)\genLt} \&= 
a(-u)\genLt 
+2\parfixP \frac{u^2-\beta^2}{\beta^2} \brk[s]*{ a'(-u) - \frac{ua(-u)}{u^2-\beta^2}}\genPt
\\
\ad(\ukmat_{\text{null}})\brk!{a(u)\genPt} \&= 
a(-u)\genPt
\\
\ad(\ukmat_{\text{null}})\brk!{\genDLt} \&= -\genDLt 
+\frac{2\beta}{u^2-\beta^2}\genLt
+ \brk[s]*{\lambda'(u)-\frac{u\lambda(u)}{u^2-\beta^2}}\genP
\\
\ad(\ukmat_{\text{null}})\brk!{\genDP} \&= -\genDP +\frac{2}{\beta} \genPt
\>
Here, $\lambda(u)$ parametrises the deformation degrees of freedom 
for the k-matrix.
The reduction only yields terms which are proportional to $u$,
but in general any odd function of the rapidity variable is permissible.
The continuous degrees of freedom of the k-matrix can be written as the deformation
\[
\ukmat_{\text{null}}
=
\ukmat_{\text{null},0}
\exp\brk!{2\parfixP\genDP+\lambda(u)\genPt}
\]
%

\paragraph{Permutation Case.}

The permutation k-matrix reduces without further ado.
The reduced adjoint action reads
\<@{eq:redadperm}
\ad(\ukmat_{\text{perm}})\brk!{a(u) \genLt}
\&= 
a(u) \genL
-2 \brk[s]*{u\frac{\nu^\labP}{\nu^\labL}+\parfixP}
\frac{u^2-\beta^2}{\beta^2}\brk[s]*{ a'(u) +\frac{ua(u)}{u^2-\beta^2} }\genPt
\\
\ad(\ukmat_{\text{perm}})\brk!{a(u)\genPt} \&= -a(u) \genPt
\\
\ad(\ukmat_{\text{perm}})\brk!{\genDLt} \&= \genDLt
+ 2\frac{\nu^\labP}{\nu^\labL}\genDP
-
\brk[s]*{ \lambda'(u) -\frac{u\lambda(u)}{u^2-\beta^2} }\genPt
\\
\ad(\ukmat_{\text{perm}})\brk{\genDP} \&= -\genDP
\>
The reduction leads to a function $\lambda(u)$ with a constant and a linear term in $u$.
For a general deformation, it can be chosen arbitrarily.
Again, the continuous parameters are generated by
\[
\ukmat_{\text{perm}}= \ukmat_{\text{perm},0} \exp\brk!{2\parfixP\genDP+\lambda(u) \genPt}
\]
%

\paragraph{Discussion.}

In conclusion, we have found three inequivalent families of
k-matrices of the reduced bialgebra which
can be obtained from k-matrices for the original bialgebra
upon contraction and reduction.

The three families of k-matrices are distinguished by their 
action on the $\genLt$ and $\genPt$ generators
as well as the rapidity variable $u$.
Two of the families flip the sign of the rapidity $u$,
the third one leaves $u$ unchanged. 
Among the former two, one family flips the signs 
of $\genLt$ and $\genPt$ whereas the other one preserves them.
The third family flips the sign of $\genPt$ while preserving the sign of $\genLt$.
Clearly, these three k-matrix families correspond to discrete transformations
which are not connected smoothly to the identity transformation.

In terms of the scattering problem for which the k-matrices serve as
reflection matrices, the three families correspond to three distinct types of discrete transformations:
In this picture, the generator $\genPt$ measures the momentum of a particle and
$u\genPt$ measures the energy of a particle.
Hence, the diagonal k-matrix family describes an ordinary spatial reflection.
The null k-matrix family describes a temporal reflection;
it therefore serves as an ordinary reflection in the mirror scattering problem
where space and time directions are interchanged.
Finally, the permutation k-matrix incorporates a simultaneous reflection in space and time.

We note that all three families of k-matrices have some degrees of freedom
which we can associate to inner automorphisms of the affine algebra applied to the k-matrices. 
Not all inner automorphisms preserve the reflection equation,
but the allowable subsets
are generated by $\lambda(u)\genP$ and $\genDP$ in all three cases.
We note that only very few of these deformations can be traced
back to the original bialgebra before contraction, twisting and reduction.
Curiously, the three k-matrix families are based on rather distinct
types of k-matrices in the original bialgebra.

\paragraph{Non-Reductive Case.}

Finally, we would like to point out that there appears to be a potential fourth 
family of k-matrices for the reduced algebra.
It completes the combinations of sign flips for the momentum and energy
in that it preserves both.
It is given by the basic reflection 
\<@{eq:knonred}
\ad(\ukmat_{\text{perm}^*,0})\brk!{a(u) \genLt}
\&= 
-a(u) \genL
-2 \frac{\nu^\labP}{\nu^\labL}
u\frac{u^2-\beta^2}{\beta^2}\brk[s]*{ a'(u) +\frac{ua(u)}{u^2-\beta^2} }\genPt
\\
\ad(\ukmat_{\text{perm}^*,0})\brk!{a(u)\genPt} \&= a(u) \genPt
\\
\ad(\ukmat_{\text{perm}^*,0})\brk{\genDLt} \&= \genDLt
+ 2\frac{\nu^\labP}{\nu^\labL}\genDP
- \frac{2\beta}{u^2-\beta^2}\genLt
+ 2\frac{\nu^\labP}{\nu^\labL}\frac{\beta}{u^2-\beta^2}\genPt
\\
\ad(\ukmat_{\text{perm}^*,0})\brk{\genDP} \&= -\genDP + \frac{2}{\beta}\genPt
\>
This k-matrix is an automorphism and it satisfies the reflection equation.

The existence of this k-matrix is curious because we cannot obtain it via reduction:
The k-matrix can be traced back to an automorphism in the contracted algebra
\<@{eq:knonredpre}
\ad(\ukmat_{\text{perm}^*,0})\brk!{a(u)\genL^+} \&= 
-a(u) \genL^+ 
-2 \frac{\nu^\labP}{\nu^\labL} ua'(u) \genP^+
\\
\ad(\ukmat_{\text{perm}^*,0})\brk!{a(u)\genL^0} \&= 
a(u) \brk[s]*{\genL^0 -\eta u \genL^+ -\frac{\nu^\labP}{\nu^\labL}\eta u\genP^+ }
+2\frac{\nu^\labP}{\nu^\labL}ua'(u)
\brk!{\genP^0-\eta u \genP^+}
\\
\ad(\ukmat_{\text{perm}^*,0})\brk!{a(u)\genL^-} \&= 
a(u) \brk[s]*{-\genL^- + 2\eta u \genL^0 -\eta^2 u^2 \genL^+
+2\frac{\nu^\labP}{\nu^\labL}\eta u \genP^0
-2\frac{\nu^\labP}{\nu^\labL}\eta^2 u^2 \genP^+
}
\\;
+2\frac{\nu^\labP}{\nu^\labL} ua'(u)
\brk!{
-\genP^-
+2\eta u \genP^0
-\eta^2 u^2 \genP^+
}
\\
\ad(\ukmat_{\text{perm}^*,0})\brk{\genDL} \&= 
\genDL + 2\frac{\nu^\labP}{\nu^\labL}\genDP 
-\eta \genL^+ - \frac{\nu^\labP}{\nu^\labL}\eta \genP^+
\\
\ad(\ukmat_{\text{perm}^*,0})\brk!{a(u)\genP^+} \&= 
a(u) \genP^+ 
\\
\ad(\ukmat_{\text{perm}^*,0})\brk!{a(u)\genP^0} \&= 
a(u) \brk!{-\genP^0 +\eta u \genP^+}
\\
\ad(\ukmat_{\text{perm}^*,0})\brk!{a(u)\genP^-} \&= 
a(u) \brk!{\genP^- - 2\eta u \genP^0 +\eta^2 u^2 \genP^+}
\\
\ad(\ukmat_{\text{perm}^*,0})\brk{\genDP} \&= -\genDP + \eta \genP^+
\>
This k-matrix reduced to the desired form upon setting the parameter $\eta=2\econst^{-\iunit\alpha}/\beta$.
Since $\genP^a$ transforms oppositely to $\genL^a$,
this case appears to be of the permutation kind. 
Furthermore, it is of the form of a null reflection.
By adding further twist terms of the form 
$\genL^+\wedge\genP^0$ and $(u\genL^+)\wedge\genP^+$ and $\genL^+\wedge(u\genP^+)$
to the r-matrix, 
we have attempted to satisfy the reflection equation
while paying attention to the Yang--Baxter equation.
However, it turns out that we cannot satisfy both equations at the same time
if also the r-matrix is to properly reduce as described in \secref{sec:reduction}.
We can merely ensure any two out of these three properties,
but this is insufficient for a full derivation of the k-matrix via reduction.

\section{Supersymmetry}
\label{sec:supersymmetry}

The construction of the Poincaré bialgebra by contraction and reduction 
can be readily extended by supersymmetry:
\[`{}@{eq:fig-susy}
\begin{mpostfig}[opt=align]
nums[1]:=1.8cm;
nums[2]:=3.5cm;
nums[3]:=4.0cm;
nums[4]:=4.0cm;
vpos[1]:=(0,0);
vpos[2]:=vpos[1]+(0.5nums[1]+0.5nums[2]+0.5cm,0);
vpos[3]:=vpos[2]+(0.5nums[2]+0.5nums[3]+0.5cm,0);
vpos[4]:=vpos[3]+(0.5nums[3]+0.5nums[4]+0.5cm,0);
paths[1]:=rbox(vpos[1],(nums[1],1cm),5pt);
paths[2]:=rbox(vpos[2],(nums[2],1.5cm),5pt);
paths[3]:=rbox(vpos[3],(nums[3],1.5cm),5pt);
paths[4]:=rbox(vpos[4],(nums[4],1.5cm),5pt);
for i:=1 upto 4: fillshape(paths[i],0.9white); endfor
label.bot(btex $\alg{sl}(2)$ etex, point 4.5 of paths[1]);
label.bot(btex $\alg{sl}(2)\times\alg{sl}(2)$ etex, point 4.5 of paths[2]);
label.bot(btex $\alg{sl}(2)\ltimes\Real^{2,1}$ etex, point 4.5 of paths[3]);
label.bot(btex $\alg{gl}(1)\ltimes\Real$ etex, point 4.5 of paths[4]);
label(btex $\alg{sl}(2)\times\alg{d}(2,1;\epsilon)$ etex, center paths[2]);
label(btex $\alg{sl}(2)\ltimes\alg{psu}(2|2)\ltimes\Real^{2,1}$ etex, center paths[3]);
label(btex $\alg{gl}(1)\ltimes\alg{psu}(2|2)\ltimes\Real$ etex, center paths[4]);
label.top(btex $\alg{d}(2,1;\epsilon)$ etex, point 0.5 of paths[1]);
label.top(btex $\alg{d}(2,1;\epsilon)\times\alg{d}(2,1;\epsilon')$ etex, point 0.5 of paths[2]);
label.top(btex $\alg{sl}(2)\ltimes\alg{psu}(2|2)^2\ltimes\Real^{2,1}$ etex, point 0.5 of paths[3]);
label.top(btex $\alg{gl}(1)\ltimes\alg{psu}(2|2)^2\ltimes\Real$ etex, point 0.5 of paths[4]);
label(btex pairing etex rotated -90, 0.5[point 2.5 of paths[1],point 6.5 of paths[2]]);
label(btex contraction etex rotated -90, 0.5[point 2.5 of paths[2],point 6.5 of paths[3]]);
label(btex reduction etex rotated -90, 0.5[point 2.5 of paths[3],point 6.5 of paths[4]]);
\end{mpostfig}
\]
The resulting supersymmetric r-matrices yield
the classical limit of the scattering matrix
for the scattering problem of AdS/CFT integrability.
In particular, this extends the kinematical discussion
of the previous section by flavour degrees of freedom.
In the following we discuss the k-matrix structures
in the supersymmetric setting.

\subsection{k-Matrices for \texorpdfstring{$\alg{d}(2,1;\epsilon)$}{d(2,1;epsilon)}}
\label{sec:susyk}

The supersymmetry extension of the above construction
replaces one of the two $\alg{sl}(2)$ factors
by the exceptional Lie superalgebra $\alg{d}(2,1;\epsilon)$. 
Let us therefore start by considering k-matrices
for the rational $\alg{d}(2,1;\epsilon)$ affine bialgebra.
The following discussion is sketchy and merely outlines
the construction; we shall not need a complete and detailed treatment
for our applications.

The even part of the $\alg{d}(2,1;\epsilon)$ superalgebra
is composed of three $\alg{sl}(2)$ algebras;
we denote their generators by $\genM^a$, $\genH^a$ and $\genK^a$.
The subalgebra $\alg{sl}(2)_\labM$ serves as one of the $\alg{sl}(2)_\labAB$ factors
of the construction outlined in \secref{sec:pairing}.
The other two subalgebras $\alg{su}(2)_{\labH,\labK}$ serve as additional flavour symmetries
in the scattering problem.
These are coupled by eight supercharges $\genQ^{j,lr}$
with $j,l,r={\uparrow},{\downarrow}$
transforming in spin-\vfrac{1}{2} representations 
of each $\alg{sl}(2)$ factor
\<
\comm{\genM^a}{\genQ^{j, l r}} 
\&= 
- \half(\slpauli^a)^j{}_k \genQ^{k, l r}
\\
\comm{\genH^a}{\genQ^{j, l r}} 
\&= 
- \half(\slpauli^a)^l{}_m \genQ^{j, m r} 
\\
\comm{\genK^a}{\genQ^{j, l r}} 
\&= 
- \half(\slpauli^a)^r{}_p \genQ^{j, l p}
\>
In our normalisation, the odd Lie brackets read
\[
\acomm{\genQ^{j,lr}}{\genQ^{k,mp}} =
-\epsilon (\slpauli_a\sleps)^{jk}\sleps^{lm}\sleps^{rp}\genM^a
-(1-\epsilon) \sleps^{jk}(\slpauli_a\sleps)^{lm}\sleps^{rp}\genH^a
+\sleps^{jk}\sleps^{lm}(\slpauli_a\sleps)^{rp}\genK^a
\]
Finally, we normalise the quadratic invariant $\genJ^2_{12}$
to be an extension of $\genM_1\sldot\genM_2$
\[
\casJ_{12}=
\genM_1\sldot\genM_2 + (\epsilon^{-1}-1)\genH_1\sldot\genH_2-\epsilon^{-1}\genK_1\sldot\genK_2-\epsilon^{-1}\casQ_{12}
\]~
with the odd quadratic combination
\[
\casQ_{12} \defeq \half \sleps_{jk} \sleps_{lm} \sleps_{rp}
\genQ^{j, lr}_1\genQ^{k, mp}_2
\]
The plain rational r-matrix $\urmat_{12}\simeq\casJ_{12}/(u_1-u_2)$
straight-forwardly generalises the $\alg{sl}(2)$ case in \eqref{eq:urrat}. 

\paragraph{k-Matrices.}

The above composition of the superalgebra
makes it particularly convenient to construct suitable k-matrices
based on the established k-matrices of $\alg{sl}(2)$:
As a (bosonic) k-matrix does not transform between
even and odd generators, it is evident that a k-matrix
must separately satisfy the reflection equation
on the even and odd parts of the r-matrix.
The even part of the r-matrix
is composed from quadratic invariants
of the three $\alg{sl}(2)$ factors,
and hence the k-matrix must act on each $\alg{sl}(2)$ subalgebra
according to the general form with 4 parameters discussed in \secref{sec:kmat}.
As the three 4-parameter families of k-matrices in \eqref{eq:ukrat}
are generated by inner automorphisms,
the action of the combined k-matrix extends canonically to the supercharges.
The odd part of the reflection equation then imposes further constraints
among the 12 parameters. Let us discuss and solve these constraints:

The $\alg{sl}(2)$ k-matrix in \eqref{eq:ukrat}
acts according to a particular fixed direction $n^a$.
The combination $\dirrefl\sldot\slgen$
generates a rotation around this direction,
and its adjoint action measures
the corresponding charge of the generator it acts upon.
For the three $\alg{sl}(2)$ generators, these charges are $0$ and $\pm 1$.
Consequently, the exponential in \eqref{eq:ukrat}
produces the prefactors $1$ or 
\[
\frac{\parrefl\pm (u-\parfix)}{\parrefl\mp (u-\parfix)}
\],
which all constitute rational functions in $u$.
These rational functions conspire to cancel in the even part of the reflection equation.

Conversely, the adjoint action of $\dirrefl\sldot\slgen$
on the supercharges has the eigenvalues $\pm{\vfrac{1}{2}}$.
The prefactor generated by the adjoint action of the k-matrix
on the supercharges are thus for each $\alg{sl}(2)$ factor:
\[
\sqrt{\frac{\parrefl\pm (u-\parfix)}{\parrefl\mp (u-\parfix)}}
\]
Quite clearly, the fractional zeros and poles cannot cancel on their own
within the reflection equation.
\unskip\footnote{Even though the case $\parrefl=0$ yields constant prefactors $\pm\iunit$,
the sign cannot be assigned consistently to let all terms cancel in the reflection equation.}
Hence, the parameters for the three $\alg{sl}(2)$ k-matrices should be adjusted
such that all zeros and poles either cancel or come with unit degree.
This is achieved by selecting two of the $\alg{sl}(2)$ factors
to support a non-trivial prefactor with matching parameters $\parrefl$
while the third $\alg{sl}(2)$ prefactor has to be trivial with $\parrefl=\infty$.
Furthermore, the fixed point $\parfix$ of the rapidity parameter reflection
should be universal.
These constraints turn out to be sufficient to satisfy the reflection equation.

In conclusion, we find three families of k-matrices
distinguished by the $\alg{sl}(2)$ factor on which it acts isotropically.
Each family has 6 parameters: one universal fixed point $\parfix$,
one universal pole parameter $\parrefl$
and two orientation unit vectors $\dirrefl^a$
accounting for two parameters each.
Starting with the basic reflection
\[
\ad(\ukmat_0)\brk!{a(u)\gen{X}} = a(-u) \gen{X}
\&
\ad(\ukmat_0)\brk{\genD} = -\genD
\],
we can write these as deformations, e.g.\ for the family linking
$\alg{sl}(2)_\labM$ and $\alg{su}(2)_\labH$ 
\[@{eq:ukratd21eps}
\ukmat_{\text{gen}}
\&=
\ukmat_0
\exp\brk!{2\parfix\genD}
\exp\brk[s]*{
\log\brk*{\frac{\parrefl+u-\parfix}{\parrefl-u+\parfix}}
\brk!{\dirrefl_\labM\sldot\genM+\dirrefl_\labH\sldot\genH}
}
\]
We point out that the above combination
also has a null limit as described in \secref{sec:diagnull}.
The deformed k-matrix takes a form extending \eqref{eq:ukratnull} to
\[
\ukmat_{\text{null}}
\&=
\ukmat_0
\exp\brk!{2\parfix\genD}
\exp\brk[s]!{2(u-\parfix)\brk!{\nullrefl_\labM\sldot\genM+\nullrefl_\labH\sldot\genH}}
\]~
with two independent null vectors $\nullrefl_\labM$ and $\nullrefl_\labH$ and a shared shift parameter $\parfix$.
\unskip\footnote{The definition of the null limit allows to assign 
different speeds for approaching the two null vectors.
Hence, the magnitudes of the two null vectors can be chosen independently.}
Altogether, this amounts to 5 independent parameters for null k-matrices.

The above exhausts the admissible k-matrices for generic values of
the algebra parameter $\epsilon$.
\unskip\footnote{In fact, $\ukmat_0$ can be combined with the automorphism
that flips the sign of all odd generators.
This k-matrix has no relevant deformations,
and it is not suited for the reduction step later on.}
For additional outer automorphisms of the superalgebra $\alg{d}(2,1;\epsilon)$, 
consider the three groups of special values,
namely $\epsilon=0,1,\infty$ and $\epsilon=-1,\half,2$ and $\epsilon=(\pm\iunit)^{2/3}$.
The latter one has automorphisms of order 3 which do not work as reflections.
The middle one represents the superalgebra $\alg{osp}(4|2)$
which may have additional k-matrices.
We are concerned with the limit $\epsilon\to0$,
hence the point $\epsilon=0$ is of interest.
Here, there exists an outer automorphism which interchanges
the subalgebras $\alg{su}(2)_\labH$ and $\alg{su}(2)_\labK$.
Correspondingly, the supercharges transform as
$\genQ^{j,lr}\to \pm\iunit \genQ^{j,rl}$.
As discussed in \secref{sec:pairing},
the even part of the reflection equation \eqref{eq:algrefeq}
requires the rapidity parameters to remain invariant under the automorphism,
$u\to +u$, see \eqref{eq:permrescale}. 
Unfortunately, the odd part is not solved by this automorphism
because $\ad(\ukmat_{\text{perm}})_1\urmat_{12}=+\ad(\ukmat_{\text{perm}})_2\urmat_{12}$
such that these two terms will not cancel in the reflection equation.

\paragraph{Null Twists.}

The reduction step relies on null twists introduced in \secref{sec:nulltwist}.
Such twists extend to the superalgebra $\alg{d}(2,1;\epsilon)$, let us discuss the available options:
An $\alg{sl}(2)$ null twist can be applied
to each of the three factors with an individual null vector $\dirtw^a$
parametrising its direction and strength.
In addition, a twist involving the supercharges 
may be applied provided that its parameters match with the twist for 
either of the $\alg{sl}(2)$ subalgebras.
For example, the combination of an odd twist with a twist for $\alg{sl}(2)_\labM$
takes the form:
\[
\dirtw^c_\labM
\brk!{\iunit f_{abc}\genM^a_1\genM^b_2
+\quarter\epsilon^{-1}(\sleps\slpauli_c)_{jk}\sleps_{lm}\sleps_{rp}\genQ^{j,lr}_1\genQ^{k,mp}_2
}
\]
Similar combined twists can be constructed for the other two $\alg{sl}(2)$ subalgebras,
but the Yang--Baxter equation excludes
the simultaneous activation of more than one combined twist.

Twisting has an imprint on the k-matrix:
The adjoint action on each $\alg{sl}(2)$ factor
has to be in accordance with the null-twisted k-matrix in \eqref{eq:ukrattw}.
Here, the rational prefactor is generically promoted to degree 2.
For a proper combination of poles and zeros on the supercharges
along the lines discussed above,
we therefore have to ensure that both the parameter $\parrefl$
and the twist-related combination $\nu^{-1}\dirtw\sldot\dirrefl$
agree between both $\alg{sl}(2)$ factors on which the k-matrix
acts non-isotropically.
For example, for the combined twist
\[
\iunit \dirtw^c_\labM f_{abc}\genM^a_1\genM^b_2
+\iunit (\epsilon^{-1}-1) \dirtw^c_\labH f_{abc}\genH^a_1\genH^b_2
\],
we have to ensure that
\[@{eq:nulltwistsusyconstraint}
\dirtw_\labM\sldot\dirrefl_\labM
=
\dirtw_\labH\sldot\dirrefl_\labH
\]
Then, the logarithm term in the k-matrix deformation equally applies to 
both $\alg{sl}(2)_\labM$ and $\alg{su}(2)_\labH$ factors
\<
\ukmat
\&=
\ukmat_0
\exp\brk!{2\parfix\genD}
\exp\brk[s]*{
\log\brk*{\frac{\parrefl+u-\parfix+\nu^{-1}\dirtw_\labM\sldot\dirrefl_\labM (u-\parfix)^2}
               {\parrefl-u+\parfix+\nu^{-1}\dirtw_\labM\sldot\dirrefl_\labM (u-\parfix)^2}}
\brk!{\dirrefl_\labM\sldot\genM+\dirrefl_\labH\sldot\genH}
}
\>
In that case, the odd part of the reflection equation requires
that there is a non-trivial odd twist
corresponding to either of the two non-isotropic $\alg{sl}(2)$ factors,
\[
\quarter\epsilon^{-1}(\sleps\slpauli\sldot\dirtw_\labM)_{jk}\sleps_{lm}\sleps_{rp}\genQ^{j,lr}_1\genQ^{k,mp}_2
\eqnjoin{\text{or}}
\quarter\epsilon^{-1}\sleps_{jk}(\sleps\slpauli\sldot\dirtw_\labH)_{lm}\sleps_{rp}\genQ^{j,lr}_1\genQ^{k,mp}_2
\]
Altogether, a null-twisted r-matrix permits a suitable k-matrix 
subject to the additional constraint \eqref{eq:nulltwistsusyconstraint}.
The null limit for the k-matrix orientation is analogous 
to the purely bosonic case in \eqref{eq:ukrattwnull}.

\subsection{Pairing}
\label{sec:susypairing}

Next, we pair the $\alg{sl}(2)_\labM$ subalgebra of $\alg{d}(2,1;\epsilon)$
with another $\alg{sl}(2)$ factor along the lines of \secref{sec:pairing}.
The latter can be a standalone $\alg{sl}(2)$
or the $\alg{sl}(2)_\labM$ subalgebra of another $\alg{d}(2,1;\epsilon')$.
In other words, we consider the combinations
\[
\alg{sl}(2)_\labA\times\alg{d}(2,1;\epsilon)_\labB
\&
\alg{d}(2,1;\epsilon)_\labA\times\alg{sl}(2)_\labB 
\&
\alg{d}(2,1;\epsilon)_\labA \times \alg{d}(2,1;\epsilon')_\labB
\]
Each of these combinations features a bosonic $\alg{sl}(2)_\labA\times \alg{sl}(2)_\labB$ subalgebra
of the kind discussed in \secref{sec:contractreduct}.
Importantly, we fuse the two bialgebra factors with
a mixed r-matrix twist of the form $(\genM_\labA^0+\genM_\labB^0)\wedge\genM_\labB^+$
as in \eqref{eq:rpair}.

\paragraph{Diagonal and Null Cases.}

First, we consider the case of a k-matrix that maps each $\alg{sl}(2)$ factor to itself.
The subalgebras $\alg{sl}(2)_\labAB$ may each be part of a superalgebra or form a standalone algebra;
k-matrices for standalone $\alg{sl}(2)$ algebras have been discussed in \secref{sec:kmat}.
If $\alg{sl}(2)$ is part of a superalgebra, 
the k-matrix must be combined with the k-matrix of another $\alg{sl}(2)$ part of this superalgebra,
as discussed in \secref{sec:susyk};
\unskip\footnote{We have already seen that an isotropic action of the k-matrix on $\genM^a$
is incompatible with the reduction step in \secref{sec:reduction},
and therefore we can disregard this case.}
without loss of generality, we may assume that it is linked to $\alg{su}(2)_\labH$.

Consequently, the k-matrices in this case are the straight-forward supersymmetric generalisation
of those obtained in \secref{sec:pairing};
more concretely:
The mixed twist and the reflection equation
imply that the k-matrix must act trivially on either $\genM_\labA^0$ or $\genM_\labB^+$.
This implies that either the k-matrix for $\alg{sl}(2)_\labA$ is diagonal
\[
\ukmat_{\text{diag}} = \ukmat^{\labA}_{\text{diag}} \ukmat^{\labB}_{\text{tw}}
\],
or the k-matrix for $\alg{sl}(2)_\labB$ is null
\[
\ukmat_{\text{null}} = \ukmat^{\labA}_{\text{gen}} \ukmat^{\labB}_{\text{null}}
\]
As far as $\alg{sl}(2)_\labAB$ are part of a superalgebra,
the supersymmetric completion of the corresponding k-matrix is described in \secref{sec:susyk}.

Altogether, we find that the k-matrices in 
\eqref{eq:ukratcontdiag} and \eqref{eq:ukratcontnull}
with six parameters
can be extended by supersymmetry.
This includes matching the automorphism on $\alg{sl}(2)_\labM$
with a similar automorphism acting on $\alg{su}(2)_\labH$
for which the orientation can be adjusted with two additional parameters.
The constraint \eqref{eq:nulltwistsusyconstraint}
is automatically satisfied for our purposes because
the reflection vectors $\dirrefl^a$
will always be chosen to be orthogonal to the null twist vectors $\dirtw^a$,
so that both sides of the constraint are zero.
In fact, we may disable the twist $\dirtw_\labH^a=0$
for the subalgebra $\alg{su}(2)_\labH$.

\paragraph{Permutation Case.}

For the combination of two superalgebras with equal parameters $\epsilon_\labA=\epsilon_\labB$
\footnote{If the parameters of the two superalgebras 
are related by an equivalence, e.g.\ $\epsilon_\labA=-\epsilon_\labB/(1-\epsilon_\labB)$,
the k-matrix automorphism should be supplemented by
the corresponding permutations among the $\alg{sl}(2)$ factors.
This case is equivalent to $\epsilon_\labA=\epsilon_\labB$
upon a relabelling of two of the $\alg{sl}(2)$ factors.}
we may construct a k-matrix family for the basic permutation automorphism
\[
\ad(\ukmat_{\text{perm},0})\brk!{a(u)\gen{X}^{\labAB}} \&= a(\sigma^{\pm1} u)\gen{X}^{\labBA}
\&
\ad(\ukmat_{\text{perm},0})\brk!{\genD^{\labAB}} \&= \sigma^{\mp1}\genD^{\labBA}
\],
where $\sigma=-\nu_\labA/\nu_\labB$ according to \eqref{eq:permrescale}.
The permutations $\ukmat_{\text{perm}}$ are constructed in analogy to \eqref{eq:ukperm},
and we find one parameter $\parfix$ for shifts and 
three sets of three parameters $\parrot_a^{\labM,\labH,\labK}$
for isomorphisms between the $\alg{sl}(2)$ factors.
In the presence of the mixed null twist, we have to constrain $\parrot_-^\labM=0$.
As the k-matrix is rapidity-independent,
there is no need to match the twist 
on the other $\alg{sl}(2)$ factors or in the supersymmetric sector.

\subsection{Poincaré Superalgebra}

The contraction limit of $\alg{d}(2,1;\epsilon)\times\alg{sl}(2)$ 
or $\alg{d}(2,1;\epsilon_\labA)\times\alg{d}(2,1;\epsilon_\labB)$ 
to a Poincaré superalgebra follows along the same lines as
described in \secref{sec:contraction}.
The two (sub)algebras $\alg{sl}(2)_{\labM,\labAB}$
are contracted as before while the superalgebra parameter(s)
$\epsilon_\labAB$ scale with the contraction parameter $\epsilon$.
In this limit, the superalgebra relations between $\genM$ and $\genQ$
change to
\[@{eq:superpoincareL}
\comm{\genL^a}{\genQ^{j, l r}} 
\&= 
- \half (\slpauli^a)^j{}_k \genQ^{k, l r}
\]~
and
\[@{eq:superpoincareQ}
\acomm{\genQ^{j,lr}}{\genQ^{k,mp}} =
-\frac{1}{\mref} (\slpauli_a\sleps)^{jk}\sleps^{lm}\sleps^{rp}\genP^a
-\sleps^{jk}(\slpauli_a\sleps)^{lm}\sleps^{rp}\genH^a
+\sleps^{jk}\sleps^{lm}(\slpauli_a\sleps)^{rp}\genK^a
\]
The limiting quadratic invariants are
\[@{eq:superpoincareJ2}
\casJ_{12}=\genL_1\sldot\genP_2+\genP_1\sldot\genL_2
+ \mref\brk!{\genH_1\sldot\genH_2-\genK_1\sldot\genK_2-\casQ_{12}}
\eqnjoin{\text{and}}
\genP_1\sldot\genP_2
\]
The r-matrix contracts into a joint structure extending \eqref{eq:urmatpoincare} to
\[@{eq:urmatsuperpoincare}
\urmat_{12}
=
-\nu_\labL\frac{\casJ_{12}}{u_1-u_2}
-\nu_\labP\frac{\genP_1\sldot\genP_2}{u_1-u_2}
+\dirtw_\labL (\genL^0\wedge\genP^+)_{12}
+\dirtw_\labP (\genP^0\wedge\genP^+)_{12}
\]

We note that the mass parameter $\mref$ now appears explicitly in above superalgebra.
However, it can be scaled to an arbitrary value by rescaling $\genP$,
and thus it carries no significance for the Poincaré superalgebra
other than defining a reference mass scale.

For the combination of two supersymmetry algebras,
we obtain two copies of the algebraic relations \eqref{eq:superpoincareL} and \eqref{eq:superpoincareQ}
with potentially different parameters $\mref_\labAB$.
For the quadratic invariant $\casJ_{12}$ in \eqref{eq:superpoincareJ2}
the combination $\genH^2-\genK^2-\genQ^2$
is doubled for the two superalgebras
with a corresponding prefactor $\mref_\labAB$.

The doubled supersymmetry algebra has two reference mass scales $\mref_\labAB$
which can be scaled simultaneously by rescaling $\genP$,
hence only their ratio matters for algebraic purposes
\[
\mratio\defeq \frac{\mref_\labA}{\mref_\labB}
\]
Furthermore, the two mass scales are exchanged by interchanging the supersymmetry algebras.
Consequently, superalgebras are equivalent if and only if
their mass ratios $\mratio$ are equal or inverse.
\unskip\footnote{Furthermore, a sign flip of $\mratio$ is equivalent
to permuting $\alg{su}(2)_\labH$ and $\alg{su}(2)_\labK$
for one of the two $\alg{psu}(2|2)$ factors.}
The fixed points under inversion $\mratio=\pm1$ are special cases 
for which the supersymmetry algebra acquires an extra $\Integer_2$-automorphism.

\paragraph{Diagonal and Null Cases.}

The analysis of the contraction limit for k-matrices
is equivalent to the one carried out in \secref{sec:contraction}.
The main difference is that the k-matrix adjoint action
now extends non-trivially to the $\alg{su}(2)_\labH$ subalgebra(s)
in order for the reflection equation to hold on the odd part of the r-matrix
as described in \secref{sec:susyk}.
Its orientation adds two parameters per superalgebra
to the untwisted and diagonal families of k-matrices ($\dirrefl_\labH^a$)
as well as to the null family ($\nullrefl_\labH^a$).

We thus find that the k-matrix family has 10 parameters in the absence of null twists
($\parfix$, $\parrefl$, $\dirrefl^a$, $\dirrefl_\labH^a$
as well as $\parfixP$, $\parreflP$, $\dirreflP^a$).
Two additional parameters describe the k-matrix orientation $\dirrefl_\labH^a$ on the second supersymmetry algebra 
when combining two supersymmetry algebras.
In the presence of the mixed twist, these parameters are constrained
according to \eqref{eq:nulltwistsusyconstraint}
in order to align the k-matrix orientation with the mixed twist;
this holds for both resulting families with diagonal and null k-matrices
(for the null k-matrix, these are $\parfix$, $\nullrefl$, $\nullrefl_\labH^a$
as well as $\parfixP$ and an unconstrained vector $\nullrefl_\labP^a$). 

We note that a construction of the k-matrix for the Poincaré superalgebra
yields the same results as taking the contraction limit of the simple superalgebras.

\paragraph{Permutation Case.}

The permutation k-matrix before contraction exists only
for the combination of two isomorphic superalgebras
with $\epsilon_\labA=\epsilon_\labB$.
The contraction leads to supersymmetry algebras
with mass ratio $\mratio=-1$,
essentially because the generator $\genP$ enters with opposite signs 
in the contraction relation \eqref{eq:contraction}.
The contraction limit then follows the same lines as before,
and we obtain a contracted k-matrix of permutation type.
The allowable deformations follow by contraction,
see \eqref{eq:ukperm} and \eqref{eq:ukpermLP}
\[
\ukmat_{\text{perm}}= \ukmat_{\text{perm},0} 
\exp\brk!{2\parfixP\genDP-2\lambda_a \genP^a
+\parrot^\labH_a (\genH^{a}_\labA-\genH^{a}_\labB)
+\parrot^\labK_a (\genK^{a}_\labA-\genK^{a}_\labB)}
\]
Altogether, we find that the permutation k-matrix has 10 parameters:
the fixed point $\parfix$ and 3 sets of 3 parameters $\lambda_a,\parrot^{\labH,\labK}_a$.
The mixed twist imposes the single constraint $\lambda_-=0$.

\paragraph{Supersymmetry Permutation.}

As discussed above, the Poincaré superalgebra with mass ratio $\mratio=\pm1$
supports a permutation automorphism which may serve as a k-matrix.
The case $\mratio=-1$ follows by contraction of two equal superalgebras
as previously discussed.
Conversely, the automorphism for two superalgebras with $\mratio=+1$
exists only within the contraction limit.
\unskip\footnote{The automorphism might
arise from an alternative formulation of the contraction limit.}
Let us therefore discuss it in detail.

The corresponding plain k-matrix is defined by the adjoint action
\<
\ad(\ukmat_{\text{perm}',0})\brk!{a(u)\genL^a} \&= a(-u)\genL^a 
\&
\ad(\ukmat_{\text{perm}',0})\brk!{\genDL} \&= -\genDL 
\\
\ad(\ukmat_{\text{perm}',0})\brk!{a(u)\genP^a} \&= a(-u) \genP^a
\&
\ad(\ukmat_{\text{perm}',0})\brk{\genDP} \&= -\genDP
\\
\ad(\ukmat_{\text{perm}',0})\brk!{a(u)\gen{X}_\labAB} \&= a(-u)\gen{X}_\labBA
\&
\gen{X}\in \genH,\genK,\genQ
\>
This k-matrix clearly satisfies the reflection equation
for an untwisted r-matrix \eqref{eq:urmatsuperpoincare}.
It differs from the previous permutation case in that 
the rapidity $u$ is flipped and that $\genL^a$ and $\genP^a$
transform in parallel.
\unskip\footnote{A conceivable alternative case
is to extend the bosonic permutation case
discussed further above with no permutation of the supersymmetric extensions.
Such an automorphism does exist, however, it is not an ordinary involution
but it has a supersymmetric $\Integer_4$-structure,
i.e.\ $\ukmat^2=(-1)^{\gen{F}}$ with $\gen{F}$ the fermion number operator.}

Next we consider the admissible deformations
which follow from and are constrained by previous considerations.
The transformation of the Poincaré sector comprising $\genL^a$ and $\genP^a$
can be inferred from \secref{sec:contraction} (excluding the permutation case):
it is governed by 
the pole location $\parreflL$,
the orientation unit vector $\dirreflL^a$ (with $\dirreflL\sldot\dirreflL=-1$),
the shift parameter $\parfix$
as well as their $\labP$ deformations
$\parreflP$, $\dirreflP^a$ (with $\dirreflL\sldot\dirreflP=0$) and $\parfixP$
amounting to 8 parameters in total.
The discussion \secref{sec:susyk} shows that the transformation of $\genL^a$
implies a corresponding transformation on the supercharges $\genQ^j$.
For generic parameters $\parreflL$,
these transformations depend on the rapidity $u$.
However, the automorphism for $\genQ^j$
permutes between the subalgebras $\labAB$,
and we have seen that permutation automorphisms
typically do not admit $u$-dependent transformations
as solutions to the reflection equation, see discussion around \eqref{eq:ukperm}.
Hence, we must restrict $\parreflL$ to $\parreflL=0$ or $\parreflL=\infty$.
In addition, we can admit a $u$-independent automorphism
for the mapping between $\genH^{a},\genK^{a}$ 
subalgebras $\labAB$ along the lines of \eqref{eq:ukperm}.
These yield another 6 deformation parameters $\parrot^{\labH,\labK}_a$.
Altogether we find the following two families of k-matrices
solving the reflection equation for an untwisted r-matrix,
c.f.\ \eqref{eq:ukperm}, \eqref{eq:ukratdiagcont} and \eqref{eq:ukratnullcont}:
\<@{eq:ukratsusy'}
\ukmat_{\text{diag}'} \&=
\ukmat_{\text{perm}',0} 
\exp\brk!{2\parfix\genDL+2\parfixP\genDP}
\\;\quad\cdot
\exp\brk!{\pm \iunit \pi \dirrefl\sldot \genL
+\dirreflP\sldot \genP
+2\parreflP(u-\parfix)^{-1}\dirrefl\sldot \genP
}
\\;\quad\cdot
\exp\brk!{\parrot^\labH\sldot (\genH_\labA-\genH_\labB)+\parrot^\labK\sldot (\genK_\labA-\genK_\labB)}
\\
\ukmat_{\text{null}'} \&=
\ukmat_{\text{perm}',0} 
\exp\brk!{2\parfix\genDL+2\parfixP\genDP}
\exp\brk!{(u-\parfix)(\nullrefl\sldot\gen{L} +\nullrefl_\labP\sldot\genP)}
\\;\quad\cdot
\exp\brk!{\parrot^\labH\sldot (\genH_\labA-\genH_\labB)+\parrot^\labK\sldot (\genK_\labA-\genK_\labB)}
\>
Note that it is crucial to interpret
the term $\exp\brk{\pm \iunit \pi \dirrefl\sldot \genL}$
within $\ukmat_{\text{diag}'}$ appropriately:
It generates a rotation about an angle of $\pm\pi$ around the 
direction $\dirrefl^a$.
For the bosonic subalgebra the sign does not matter,
but it does have an effect on the supersymmetry generators $\genQ^j$.
In order for $\ukmat_{\text{diag}'}$ to generate a $\Integer_2$-automorphism,
$\exp\brk{\pm \iunit \pi \dirrefl\sldot \genL}$
must act with opposite signs on the supersymmetry generators $\genQ^j$
belonging to the opposite subalgebras $\labAB$.

Finally, we turn on the twist degrees $\dirtw_{\labL,\labP}$ in the r-matrix.
This is fully equivalent to the discussion in \secref{sec:contraction}
because the adjustments due to the twist merely affect the Poincaré subalgebra
and not the supersymmetry extensions.
We merely have to impose the restrictions $\parrefl=0$ and $\nullrefl=0$
for the diagonal and null cases, respectively,
in order to allow the extension to supersymmetry as discussed above.
However, we observe that the restriction $\nullrefl=0$ for the null case 
imposes an obstruction towards a subsequent reduction 
which will instead require $\nullrefl=-2\econst^{-\iunit\alpha}/\beta$.
Interestingly, this incompatibility can be circumvented by 
adding an odd twist, see \secref{sec:susyk}
\unskip\footnote{The odd twist may as well be applied to the 
second supersymmetry algebra or to both.
Applying it to both requires a doubled parameter $\nullrefl=-4\dirtw^\labL/\nu^\labL$.}
\[
\dirtw_\labL
\brk{\genL^0\wedge\genP^+}_{12}
\to
\dirtw_\labL
\brk{\genL^0\wedge\genP^+}_{12}
+\iunit\mref\dirtw_\labL\sleps_{lm}\sleps_{rp}\genQ^{\uparrow,lr}_{\labA,1}\genQ^{\uparrow,mp}_{\labA,2}
\]
The contribution from the odd twist to the reflection equation
is compensated by setting $\nullrefl=-2\dirtw^\labL/\nu^\labL$
in the null reflection \eqref{eq:ukratnullcont}
\<
\ukmat_{\text{null}'} \&=
\ukmat_{\text{perm}',0} 
\exp\brk!{2\parfix\genDL+2\parfixP\genDP}
\\;\quad\cdot
\exp\brk[s]*{(u-\parfix)\brk!{
  -2\nu_\labL^{-1}\dirtw^\labL \genL^+ +\nullrefl_\labP\sldot\genP}
  +2\nu_\labL^{-1}\dirtw^\labL \parfixP\genP^+}
\\;\quad\cdot
\exp\brk!{\parrot^\labH\sldot (\genH_\labA-\genH_\labB)+\parrot^\labK\sldot (\genK_\labA-\genK_\labB)}
\>
%

\subsection{Reduction}

We can now discuss the reduction in the diagonal, null and permutation cases
in analogy to \secref{sec:reduction}.

In the diagonal case, we first have to fix the null twist
and the orientation of the reflection for $\alg{su}(2)_\labH$.
Since $\dirtw_\labM\sldot\dirrefl_\labM=0$,
the constraint \eqref{eq:nulltwistsusyconstraint}
is most conveniently satisfied by dropping
the null twist $\dirtw_\labH^a=0$.
We can then set $\dirrefl_\labH^a=\delta^{a,0}$
without loss of generality.
The adjoint action \eqref{eq:redaddiag}
is supplemented by 
\<
\ad(\ukmat_{\text{diag}})\brk!{a(u)\genH^0} \&= a(-u)\genH^0
\\
\ad(\ukmat_{\text{diag}})\brk!{a(u)\genH^\pm} \&= -a(-u)\genH^\pm
\\
\ad(\ukmat_{\text{diag}})\brk!{a(u)\genK^a} \&= a(-u)\genK^a
\\
\ad(\ukmat_{\text{diag}})\brk!{a(u)\genQ^{j,lr}} \&= (-1)^{\delta^{j,l}} a(-u)\genQ^{j,lr}
\>
Evidently, this k-matrix breaks the $\alg{psu}(2|2)$ superalgebra (at level zero)
to $\alg{sl}(1|2)$. As we have seen earlier, this k-matrix 
flips the momentum $\genP$ and preserves the energy $\genP^0$.
This k-matrix clearly corresponds to the reflection matrix
for the diagonal or $Y=0$ giant graviton or D3-D7 brane setup discussed in 
\cite{Hofman:2007xp,Correa:2008av}, 
see \cite{Regelskis:2012voq} for an overview of boundary scattering 
in the context of AdS/CFT integrability.

In the null case, we may drop the null twist and null reflection 
for $\alg{su}(2)_\labH$.
The adjoint action \eqref{eq:redadnull}
is then supplemented by 
\unskip\footnote{With a non-trivial null reflection 
the resulting expressions break $\alg{su}(2)_\labH$
and become more elaborate.}
\<
\ad(\ukmat_{\text{null}})\brk!{a(u)\genH^a} \&= a(-u)\genH^a
\\
\ad(\ukmat_{\text{null}})\brk!{a(u)\genK^a} \&= a(-u)\genK^a
\\
\ad(\ukmat_{\text{null}})\brk!{a(u)\genQ^{\downarrow,lr}} \&= 
a(-u)\genQ^{\downarrow,lr}
-2\iunit \econst^{-\iunit\alpha} \beta^{-1} u a(-u)\genQ^{\uparrow,lr}
\\
\ad(\ukmat_{\text{null}})\brk!{a(u)\genQ^{\uparrow,lr}} \&= 
a(-u)\genQ^{\uparrow,lr}
\>
This case preserves both $\alg{su}(2)_{\labH,\labK}$ as well as half the supersymmetry,
which anti-commutes onto $\genP^+$ alone
and thus leads to a non-simple superalgebra $\alg{sl}(2)^2\ltimes \Real^{0|4}$.
The other half of supersymmetry is part of the Jordan block for eigenvalue $1$,
and it gets shifted by a level-one supersymmetry.
Here, the momentum $\genP$ is preserved, but the energy $\genP^0$ is flipped.
It is not evident to which boundary configuration within AdS/CFT integrability
this k-matrix corresponds to. 
As the magnon energy rather than the magnon momentum is flipped,
it appears to correspond to a temporal boundary or likewise a mirror theory setup
where energy and momentum are exchanged. 

For the straight permutation case, the k-matrix evidently permutes the two
$\alg{psu}(2|2)$ factors. It therefore preserves the diagonal $\alg{psu}(2|2)$.
As previously discussed, the momentum $\genP$ and the energy $\genP^0$ are flipped.
The generators related to the two supersymmetry subalgebras $\labAB$
are simply exchanged.
Altogether, the transformations \eqref{eq:redadperm} are supplemented by
\[
\ad(\ukmat_{\text{perm}})\brk!{a(u)\gen{X}_\labAB} \&= a(u)\gen{X}_\labBA
\&
\gen{X}=\genH,\genK,\genQ
\]

The supersymmetry permutation case splits further into a diagonal and a null subcase.
The k-matrix transformations for the Poincaré subalgebra
are precisely the same as in \eqref{eq:redaddiag} and \eqref{eq:redadnull},
provided that the r-matrix for the null case 
has an additional odd twist as detailed in the previous section.
In the diagonal case, the adjoint action is supplemented by 
\<
\ad(\ukmat_{\text{diag}'})\brk!{a(u)\gen{X}_\labAB} \&= a(-u)\gen{X}_\labBA
\eqnsep
\gen{X}=\genH,\genK
\\
\ad(\ukmat_{\text{diag}'})\brk!{a(u)\genQ^{j,lr}_\labAB} \&= 
\pm \iunit (\delta^{j,\uparrow}-\delta^{j,\downarrow})a(-u)\genQ^{j,lr}_\labBA
\>,
where the sign for the transformation of supercharges correlates
with the subalgebra $\labAB$.
In the null case, the adjoint action is supplemented by 
\<
\ad(\ukmat_{\text{diag}'})\brk!{a(u)\gen{X}_\labAB} \&= a(-u)\gen{X}_\labBA
\eqnsep
\gen{X}=\genH,\genK
\\
\ad(\ukmat_{\text{null}'})\brk!{a(u)\genQ_\labAB^{j,lr}} \&= 
a(-u)\brk[s]!{
\genQ_\labBA^{j,lr}
-2\iunit\econst^{-\iunit\alpha}\beta^{-1} u \delta^{j,\downarrow}\genQ_\labBA^{\uparrow,lr}
}
\>
Altogether, the k-matrix for the supersymmetry permutation case
permutes the two $\alg{psu}(2|2)$ factors,
and it either flips the momentum $\genP$ (diagonal case) or the
energy $\genP^0$ (null case) while preserving the other, respectively.
The diagonal case corresponds to the horizontal
D5 brane setup \cite{Correa:2008av,Correa:2011nz}, see \cite{Regelskis:2012voq};
the null case flipping the energy rather than the momentum 
should correspond to a mirror version of it.

Finally, we comment on the non-reductive permutation case
we found in the bosonic setting. 
The k-matrix \eqref{eq:knonred}
and the automorphism \eqref{eq:knonredpre}
can be extended by supersymmetry with mass ratio $\mratio=-1$ as follows:
\<
\ad(\ukmat_{\text{perm}^*,0})\brk!{a(u) \gen{X}}_\labAB \&= a(u) \gen{X}_\labBA
\eqnsep
\gen{X}=\genH,\genK
\\
\ad(\ukmat_{\text{perm}^*,0})\brk!{a(u) \genQ^{j,lr}}_\labAB \&= 
\pm \iunit a(u)
\brk[s]!{(\delta^{j,\uparrow}-\delta^{j,\downarrow}) \genQ^{j,lr}_\labBA
+2\iunit\econst^{-\iunit\alpha}\beta^{-1} u \delta^{j,\downarrow}\genQ^{\uparrow,lr}_\labBA}
\>
In particular, this k-matrix 
preserves both the energy and momentum.

\section{Conclusions and Outlook}
\label{sec:conclusions}

In this work, we have applied the method of contraction and reduction to
derive several families of classical reflection matrices, called k-matrices,
for specific bosonic and supersymmetric symmetry algebras.
In the bosonic case, we have started with two copies of the affine $\alg{sl}(2)$ algebra
with rational r-matrices and some Drinfeld twist.
This quasi-triangular algebra admits two families of k-matrices,
one that acts individually on each $\alg{sl}(2)$ factor 
and one that permutes them. 
The former family splits further in the presence of the mixed Drinfeld twist.
Upon contraction, we obtain three families of k-matrices
for the affine 3D Poincaré algebra.
Reduction finally reduces the algebra to an unconventional affine algebra
based on one rotation and one orthogonal translation.
The corresponding k-matrices reduce without further ado,
but only for particular parameter values.

The three resulting k-matrix families can be distinguished by their action 
on particle states which are labelled by their momentum and energy.
The first family corresponds to a standard reflection which 
flips the momentum but preserves the energy.
The second family instead flips the energy but preserves the momentum.
This map can be associated to a time-like reflection,
which in the integrable 2D QFT framework can be interpreted as
a regular reflection in the mirror model which interchanges
the notion of space and time,
see also \cite{Coronado:2025xwk}.
The third family flips both momentum and energy,
and its interpretation or purpose remains obscure.
Algebraically, the three k-matrix families 
are distinguished by their action on the translation generator $\genP$ 
and on the formal variable $u$ of the loop/affine algebra.
The former two families flip the sign of $u$, and the first one flips
the sign of $\genP$ while the second one does not.
The latter family flips $\genP$ but preserves $u$.

The extension by supersymmetry leads to a similar but slightly extended outcome.
When starting with the combination of $\alg{sl}(2)$ and $\alg{d}(2,1;\epsilon)$
contraction and reduction yields one copy of $\alg{psu}(2|2)$. 
The former two families of k-matrices extend straight-forwardly. 
The momentum-flipping family breaks the residual symmetry to $\alg{su}(1|2)$,
and the energy-flipping family preserves a non-simple $\alg{su}(2)^2\ltimes\Real^{0|4}$ superalgebra. 
The third family of k-matrices would permute the factors of the algebra before contraction,
but these are inequivalent in this case and therefore no such automorphism exists.
When starting with two factors of $\alg{d}(2,1;\epsilon)$,
we obtain two factors of $\alg{psu}(2|2)$ after contraction and reduction.
All three families of k-matrices extend in this case.
For the former two families, there is an additional choice
(potentially only after contraction) to preserve or permute
the two $\alg{psu}(2|2)$ algebra factors.
We summarise the different cases of resulting k-matrices in the following table:
\[~
\begin{array}{l|rrcrrr|cl}
\kmat & 
p & e & x & u & \genL & \genP
& \alg{psu}(2|2) & \alg{psu}(2|2)^2
\\ \hline\hline
\kmat_{\text{diag}} & 
-p & +e & -x & -u & -\genL & -\genP
& \alg{su}(1|2) & \<*+l[ldelim=\{] \alg{su}(1|2)^2 \\ \alg{psu}(2|2)_{\text{diag}}\ (\mratio=+1)\>
\\ \hline
\kmat_{\text{null}} & 
+p & -e & -1/x & -u & +\genL & +\genP
& \alg{sl}(2)^2\ltimes \Real^{0|4} & \<*+l[ldelim=\{] \alg{sl}(2)^4\ltimes \Real^{0|8} \\ \alg{psu}(2|2)_{\text{diag}}\ (\mratio=+1) \>
\\ \hline
\kmat_{\text{perm}} & 
-p & -e & +1/x & +u & +\genL & -\genP
&\multicolumn{1}{c}{\text{--}}  & \alg{psu}(2|2)_{\text{diag}}\ (\mratio=-1)
\\ \hline
\kmat_{\text{perm}^*} & 
+p & +e & +x & +u & -\genL & +\genP
&\multicolumn{1}{c}{\text{--}}  & \alg{psu}(2|2)_{\text{diag}}\ (\mratio=-1)
\end{array}
\]
Each line describes the action of a k-matrix on characteristic quantities
of the scattering problem, elements the affine algebra
as well as the resulting invariant subalgebra:
In particular, $p$ and $e$ denote the momentum and energy of excitations,
$u\sim e/p$ denotes the rapidity variable and 
$x$ is commonly used to describe the fundamental evaluation representation
for the supersymmetric algebra \cite{Beisert:2007ty} obeying the relations
$u=x+1/x$, $p\sim1/(x-1/x)$, $e\sim (x+1/x)/(x-1/x)$.

Our construction can be extended in several ways.
Here, we focused on the case of rational r-matrices and corresponding k-matrices. 
The contraction and reduction method extends straight-forwardly to trigonometric r-matrices.
The family of trigonometric k-matrices for the simple algebras at the starting point
has a form analogous to their rational counterparts.
Hence, we expect that contraction and reduction extends to k-matrices equally well.

The remaining step for completion of the algebraic derivation of scattering matrices
for the context of AdS/CFT integrability is to promote the contraction and reduction
to quantum algebra. Initial steps in this directions are taken in \cite{Beisert:2026xxx},
but the application to the universal R-matrix remains a challenge. 
The extension to boundary scattering would involve a universal K-matrix
\cite{2015arXiv150706276B,Appel:2020zze,Appel:2022scp},
which is an automorphism element for the quantum algebra satisfying the quantum boundary Yang--Baxter equation.
With the correct choice of parameters, this universal K-matrix should contract and reduce properly.

Another step forward would be to extend the reflection matrix by 
boundary degrees of freedom. Such boundary degrees of freedom transform
in the invariant subalgebra under reflection, and the reflection matrix
becomes a two-site operator, with the bulk site acting on the full algebra
and the boundary site acting on the subalgebra.
\unskip\footnote{The presence of boundary degrees of freedom complicates the setup considerably:
Suppose the quantum K-matrix expands as $\Kmat(u)=\kmat(u)+\hbar \kmat'(u)+\ldots$.
Even though boundary scattering is governed by $\kmat(u)$ alone,
and there is no clear physical meaning to $\kmat'(u)$,
the classical limit of the reflection equation
does receive additional contributions of the form
$\comm{\kmat'_1(u_1)}{\kmat_2(u_2)} + \comm{\kmat_1(u_1)}{\kmat'_2(u_2)}$.
These may be viewed to relax the equation by allowing
residual terms in the image of the commutator with $\kmat(u)$.}

\pdfbookmark[1]{Acknowledgements}{ack}
\subsection*{Acknowledgements}

We thank 
Timon Leupp,
Florian Loebbert
and
Mario Morales Gonzalez
for discussions.
The work of NB was supported by the Swiss National Science Foundation 
through the NCCR SwissMAP.

\ifnum\version<\verarxiv\relax
\begin{bibtex}[\jobname]

@article{Beisert:2007ty,
  author = "Beisert, Niklas and Spill, Fabian",
  title = "The Classical r-matrix of AdS/CFT and its Lie Bialgebra Structure",
  eprint = "0708.1762",
  archivePrefix = "arXiv",
  primaryClass = "hep-th",
  reportNumber = "AEI-2007-116, HU-EP-07-31",
  doi = "10.1007/s00220-008-0578-2",
  journal = "Commun. Math. Phys.",
  volume = "285",
  pages = "537--565",
  year = "2009"
}

@ARTICLE{2015arXiv150706276B,
       author = {{Balagovic}, Martina and {Kolb}, Stefan},
        title = "{Universal K-matrix for quantum symmetric pairs}",
         year = 2015,
          doi = {10.48550/arXiv.1507.06276},
archivePrefix = {arXiv},
       eprint = {1507.06276},
 primaryClass = {math.QA},
}

@ARTICLE{2014arXiv1405.5506S,
       author = {{Schrader}, Gus},
        title = "{Integrable systems from the classical reflection equation}",
         year = 2014,
          doi = {10.48550/arXiv.1405.5506},
archivePrefix = {arXiv},
       eprint = {1405.5506},
 primaryClass = {math-ph},
}

@article{Appel:2020zze,
    author = "Appel, Andrea and Vlaar, Bart",
    title = "{Universal K-matrices for quantum Kac-Moody algebras}",
    eprint = "2007.09218",
    archivePrefix = "arXiv",
    primaryClass = "math.RT",
    doi = "10.1090/ert/623",
    journal = "Represent. Theory",
    volume = "26",
    number = "26",
    pages = "764--824",
    year = "2022"
}

@article{Appel:2022scp,
    author = "Appel, Andrea and Vlaar, Bart",
    title = "{Trigonometric K-matrices for finite-dimensional representations of quantum affine algebras}",
    eprint = "2203.16503",
    archivePrefix = "arXiv",
    primaryClass = "math.RT",
    doi = "10.4171/JEMS/1686",
    month = "3",
    year = "2022"
}

@article{Kulish:1992qb,
    author = "Kulish, P. P. and Sklyanin, E. K.",
    title = "{Algebraic structures related to the reflection equations}",
    eprint = "hep-th/9209054",
    archivePrefix = "arXiv",
    reportNumber = "YITP-K-980, YITP--980",
    doi = "10.1088/0305-4470/25/22/022",
    journal = "J. Phys. A",
    volume = "25",
    pages = "5963--5976",
    year = "1992"
}

@article{Cherednik:1984vvp,
    author = "Cherednik, I. V.",
    title = "{Factorizing Particles on a Half Line and Root Systems}",
    doi = "10.1007/BF01038545",
    journal = "Theor. Math. Phys.",
    volume = "61",
    pages = "977--983",
    year = "1984"
}

@article{Sklyanin:1988yz,
    author = "Sklyanin, E. K.",
    title = "{Boundary Conditions for Integrable Quantum Systems}",
    reportNumber = "E-11-86",
    doi = "10.1088/0305-4470/21/10/015",
    journal = "J. Phys. A",
    volume = "21",
    pages = "2375--2389",
    year = "1988"
}

@article{Cremmer:1990fq,
    author = "Cremmer, Eugene and Gervais, Jean-Loup",
    title = "{The Quantum strip: Liouville theory for open strings}",
    reportNumber = "LPTENS-90-32",
    doi = "10.1007/BF02101093",
    journal = "Commun. Math. Phys.",
    volume = "144",
    pages = "279--302",
    year = "1992"
}

@article{Kulish:1992ep,
    author = "Kulish, P. P. and Sasaki, R. and Schwiebert, C.",
    title = "{Constant solutions of reflection equations and quantum groups}",
    eprint = "hep-th/9205039",
    archivePrefix = "arXiv",
    reportNumber = "YITP-U-92-07",
    doi = "10.1063/1.530382",
    journal = "J. Math. Phys.",
    volume = "34",
    pages = "286--304",
    year = "1993"
}

@article{Coronado:2025xwk,
    author = "Coronado, Frank and Komatsu, Shota and Zarembo, Konstantin",
    title = "{Coulomb branch and integrability}",
    eprint = "2506.07222",
    archivePrefix = "arXiv",
    primaryClass = "hep-th",
    doi = "10.1007/JHEP10(2025)143",
    journal = "JHEP",
    volume = "10",
    pages = "143",
    year = "2025"
}

@article{MacKay:2011av,
    author = "MacKay, Niall and Regelskis, Vidas",
    title = "{Reflection algebra, Yangian symmetry and bound-states in AdS/CFT}",
    eprint = "1101.6062",
    archivePrefix = "arXiv",
    primaryClass = "hep-th",
    doi = "10.1007/JHEP01(2012)134",
    journal = "JHEP",
    volume = "01",
    pages = "134",
    year = "2012"
}

@article{Nepomechie:2011nz,
    author = "Nepomechie, Rafael I.",
    title = "{Revisiting the Y=0 open spin chain at one loop}",
    eprint = "1109.4366",
    archivePrefix = "arXiv",
    primaryClass = "hep-th",
    reportNumber = "UMTG-272",
    doi = "10.1007/JHEP11(2011)069",
    journal = "JHEP",
    volume = "11",
    pages = "069",
    year = "2011"
}

@article{Bajnok:2012xc,
    author = "Bajnok, Zoltan and Nepomechie, Rafael I. and Palla, Laszlo and Suzuki, Ryo",
    title = "{Y-system for Y=0 brane in planar AdS/CFT}",
    eprint = "1205.2060",
    archivePrefix = "arXiv",
    primaryClass = "hep-th",
    reportNumber = "UMTG-274, ITP-UU-12-15, SPIN-12-13, ITP-BUDAPEST-REPORT-660",
    doi = "10.1007/JHEP08(2012)149",
    journal = "JHEP",
    volume = "08",
    pages = "149",
    year = "2012"
}

@article{Regelskis:2012voq,
    author = "Regelskis, Vidas",
    title = "{Quantum Algebras and Integrable Boundaries in AdS/CFT}",
    year = "2012",
    note = "PhD thesis, Dept. Math., York U., U.K."
}

@article{Correa:2013em,
    author = "Correa, Diego H. and Schaposnik Massolo, Fidel I.",
    title = "{D5-brane boundary reflection factors}",
    eprint = "1301.3412",
    archivePrefix = "arXiv",
    primaryClass = "hep-th",
    doi = "10.1007/JHEP05(2013)095",
    journal = "JHEP",
    volume = "05",
    pages = "095",
    year = "2013"
}

@article{Bajnok:2013sza,
    author = "Bajnok, Zoltan and Kim, Minkyoo and Palla, Laszlo",
    title = "{Spectral curve for open strings attached to the Y=0 brane}",
    eprint = "1311.7280",
    archivePrefix = "arXiv",
    primaryClass = "hep-th",
    reportNumber = "ITP-BUDAPEST-REPORT-664",
    doi = "10.1007/JHEP04(2014)035",
    journal = "JHEP",
    volume = "04",
    pages = "035",
    year = "2014"
}

@article{Gombor:2015kdu,
    author = "Gombor, Tamas and Palla, Laszlo",
    title = "{Algebraic Bethe Ansatz for O(2N) sigma models with integrable diagonal boundaries}",
    eprint = "1511.03107",
    archivePrefix = "arXiv",
    primaryClass = "hep-th",
    reportNumber = "ITP-BUDAPEST-REPORT-667, ITP-BUDAPEST-667",
    doi = "10.1007/JHEP02(2016)158",
    journal = "JHEP",
    volume = "02",
    pages = "158",
    year = "2016"
}

@article{Ahn:2010xa,
    author = "Ahn, Changrim and Nepomechie, Rafael I.",
    title = "{Yangian symmetry and bound states in AdS/CFT boundary scattering}",
    eprint = "1003.3361",
    archivePrefix = "arXiv",
    primaryClass = "hep-th",
    reportNumber = "UMTG-266",
    doi = "10.1007/JHEP05(2010)016",
    journal = "JHEP",
    volume = "05",
    pages = "016",
    year = "2010"
}

@article{Correa:2011nz,
    author = "Correa, Diego H. and Regelskis, Vidas and Young, Charles A. S.",
    title = "{Integrable achiral D5-brane reflections and asymptotic Bethe equations}",
    eprint = "1105.3707",
    archivePrefix = "arXiv",
    primaryClass = "hep-th",
    doi = "10.1088/1751-8113/44/32/325403",
    journal = "J. Phys. A",
    volume = "44",
    pages = "325403",
    year = "2011"
}

@article{Correa:2009dm,
    author = "Correa, D. H. and Young, C. A. S.",
    title = "{Asymptotic Bethe equations for open boundaries in planar AdS/CFT}",
    eprint = "0912.0627",
    archivePrefix = "arXiv",
    primaryClass = "hep-th",
    reportNumber = "DAMTP-2009-83, YITP-09-99",
    doi = "10.1088/1751-8113/43/14/145401",
    journal = "J. Phys. A",
    volume = "43",
    pages = "145401",
    year = "2010"
}

@article{Correa:2008av,
    author = "Correa, D. H. and Young, C. A. S.",
    title = "{Reflecting magnons from D7 and D5 branes}",
    eprint = "0808.0452",
    archivePrefix = "arXiv",
    primaryClass = "hep-th",
    reportNumber = "DAMTP-08-64, DCPT-08-45",
    doi = "10.1088/1751-8113/41/45/455401",
    journal = "J. Phys. A",
    volume = "41",
    pages = "455401",
    year = "2008"
}

@article{Hofman:2007xp,
    author = "Hofman, Diego M. and Maldacena, Juan Martin",
    title = "{Reflecting magnons}",
    eprint = "0708.2272",
    archivePrefix = "arXiv",
    primaryClass = "hep-th",
    doi = "10.1088/1126-6708/2007/11/063",
    journal = "JHEP",
    volume = "11",
    pages = "063",
    year = "2007"
}

@article{Beisert:2022vnc,
    author = "Beisert, Niklas and Im, Egor",
    title = "{Classical Lie Bialgebras for AdS/CFT Integrability by Contraction and Reduction}",
    eprint = "2210.11150",
    archivePrefix = "arXiv",
    primaryClass = "hep-th",
    doi = "10.21468/SciPostPhys.14.6.157",
    journal = "SciPost Phys.",
    volume = "14",
    pages = "157",
    year = "2023"
}

@article{Beisert:2026xxx,
    author = "Beisert, Niklas and Im, Egor",
    title = "{Contraction and Reduction for Quantum Algebras}",
    note = "work in progress",
    year = "2026"
}

@article{Beisert:2024rke,
    author = "Beisert, Niklas and Im, Egor",
    title = "{Affine Classical Lie Bialgebras for AdS/CFT Integrability}",
    eprint = "2401.10327",
    archivePrefix = "arXiv",
    primaryClass = "hep-th",
    month = "1",
    year = "2024"
}

@ARTICLE{1995q.alg.....6005E,
       author = {{Etingof}, Pavel and {Kazhdan}, David},
        title = "{Quantization of Lie bialgebras, I}",
      journal = {Selecta Mathematica},
         year = 1996,
       volume = 2,
        pages = {1-41},
          doi = {10.1007/BF01587938},
archivePrefix = {arXiv},
       eprint = {q-alg/9506005},
 primaryClass = {math.QA}
}

@ARTICLE{1997q.alg.....1038E,
       author = {{Etingof}, Pavel and {Kazhdan}, David},
        title = "{Quantization of Lie bialgebras, II}",
      journal = {Selecta Mathematica},
         year = 1998,
       volume = 4,
        pages = {213},
          doi = {10.1007/s000290050030},
archivePrefix = {arXiv},
       eprint = {q-alg/9701038},
 primaryClass = {math.QA}
}

@ARTICLE{1996q.alg....10030E,
       author = {{Etingof}, Pavel and {Kazhdan}, David},
        title = "{Quantization of Lie bialgebras, III}",
      journal = {Selecta Mathematica},
         year = 1998,
       volume = 4,
        pages = {233},
          doi = {10.1007/s000290050031},
archivePrefix = {arXiv},
       eprint = {q-alg/9610030},
 primaryClass = {math.QA}
}

@article{stolin1991,
       author = {A. Stolin},
        title = "{On Rational Solutions of Yang-Baxter Equation for sl(n)}",
      journal = {Mathematica Scandinavica},
         year = 1991,
       volume = 69,
        pages = {57},
          url = {https://www.jstor.org/stable/24492600}
}

@article{Beisert:2010jr,
    author = "Beisert, Niklas and others",
    title = "{Review of AdS/CFT Integrability: An Overview}",
    eprint = "1012.3982",
    archivePrefix = "arXiv",
    primaryClass = "hep-th",
    reportNumber = "AEI-2010-175, CERN-PH-TH-2010-306, HU-EP-10-87, HU-MATH-2010-22, KCL-MTH-10-10, UMTG-270, UUITP-41-10",
    doi = "10.1007/s11005-011-0529-2",
    journal = "Lett. Math. Phys.",
    volume = "99",
    pages = "3--32",
    year = "2012"
}

@Book{Essler:2005aa,
     author    = "Essler, F. H. L. and Frahm, H. and Göhmann, F. and Klümper, A. and Korepin, V. E.",
     title     = "The one-dimensional Hubbard model",
     address   = "Cambridge, UK",
     year      = "2005",
     pages     = "690",
     publisher = "Cambridge University Press"
}

@Article{Shastry:1986bb,
    author = "Shastry, B. Sriram",
    title     = "Exact Integrability of the One-Dimensional Hubbard Model",
    journal   = "Phys. Rev. Lett.",
    volume    = "56",
    year      = "1986",
    pages     = "2453-2455",
    doi = "10.1103/PhysRevLett.56.2453",
}

@article{Beisert:2006qh,
    author = "Beisert, Niklas",
    title = "{The Analytic Bethe Ansatz for a Chain with Centrally Extended su(2$/$2) Symmetry}",
    eprint = "nlin/0610017",
    archivePrefix = "arXiv",
    reportNumber = "AEI-2006-074, PUTP-2211",
    doi = "10.1088/1742-5468/2007/01/P01017",
    journal = "J. Stat. Mech.",
    volume = "0701",
    pages = "P01017",
    year = "2007"
}

@article{Staudacher:2004tk,
    author = "Staudacher, Matthias",
    title = "{The Factorized S-matrix of CFT/AdS}",
    eprint = "hep-th/0412188",
    archivePrefix = "arXiv",
    reportNumber = "AEI-2004-107, NSF-KITP-04-122",
    doi = "10.1088/1126-6708/2005/05/054",
    journal = "JHEP",
    volume = "05",
    pages = "054",
    year = "2005"
}

@article{Beisert:2005tm,
    author = "Beisert, Niklas",
    title = "{The SU(2$/$2) dynamic S-matrix}",
    eprint = "hep-th/0511082",
    archivePrefix = "arXiv",
    reportNumber = "PUTP-2181, NSF-KITP-05-92",
    doi = "10.4310/ATMP.2008.v12.n5.a1",
    journal = "Adv. Theor. Math. Phys.",
    volume = "12",
    pages = "945--979",
    year = "2008"
}

@article{Gomez:2006va,
    author = "Gómez, César and Hernández, Rafael",
    title = "{The Magnon kinematics of the AdS/CFT correspondence}",
    eprint = "hep-th/0608029",
    archivePrefix = "arXiv",
    reportNumber = "CERN-PH-TH-2006-140, IFT-UAM-CSIC-06-37",
    doi = "10.1088/1126-6708/2006/11/021",
    journal = "JHEP",
    volume = "11",
    pages = "021",
    year = "2006"
}

@article{Plefka:2006ze,
    author = "Plefka, Jan and Spill, Fabian and Torrielli, Alessandro",
    title = "{On the Hopf algebra structure of the AdS/CFT S-matrix}",
    eprint = "hep-th/0608038",
    archivePrefix = "arXiv",
    reportNumber = "HU-EP-06-22",
    doi = "10.1103/PhysRevD.74.066008",
    journal = "Phys. Rev. D",
    volume = "74",
    pages = "066008",
    year = "2006"
}

@article{Beisert:2006fmy,
    author = "Beisert, Niklas",
    editor = "Faddeev, L. and Henneaux, M. and Kashaev, R. and Volkov, A. and Lambert, F.",
    title = "{The S-matrix of AdS/CFT and Yangian symmetry}",
    eprint = "0704.0400",
    archivePrefix = "arXiv",
    primaryClass = "nlin.SI",
    reportNumber = "AEI-2007-019",
    doi = "10.22323/1.038.0002",
    journal = "PoS",
    volume = "SOLVAY",
    pages = "002",
    year = "2006"
}

@article{Klose:2006zd,
    author = "Klose, Thomas and McLoughlin, Tristan and Roiban, Radu and Zarembo, Konstantin",
    title = "{Worldsheet scattering in AdS$_5$ $\times$ S$^5$}",
    eprint = "hep-th/0611169",
    archivePrefix = "arXiv",
    reportNumber = "ITEP-TH-61-06, UUITP-15-06",
    doi = "10.1088/1126-6708/2007/03/094",
    journal = "JHEP",
    volume = "03",
    pages = "094",
    year = "2007"
}

@article{Torrielli:2007mc,
    author = "Torrielli, Alessandro",
    title = "{Classical r-matrix of the su(2$/$2) SYM spin-chain}",
    eprint = "hep-th/0701281",
    archivePrefix = "arXiv",
    reportNumber = "MIT-CTP-3809",
    doi = "10.1103/PhysRevD.75.105020",
    journal = "Phys. Rev. D",
    volume = "75",
    pages = "105020",
    year = "2007"
}

@article{Beisert:2017xqx,
    author = "Beisert, Niklas and Hecht, Reimar and Hoare, Ben",
    title = "{Maximally extended sl(2$/$2), q-deformed d(2,1;$\epsilon$) and 3D kappa-Poincaré}",
    eprint = "1704.05093",
    archivePrefix = "arXiv",
    primaryClass = "math-ph",
    doi = "10.1088/1751-8121/aa7a2f",
    journal = "J. Phys. A",
    volume = "50",
    number = "31",
    pages = "314003",
    year = "2017"
}

@article{Moriyama:2007jt,
    author = "Moriyama, Sanefumi and Torrielli, Alessandro",
    title = "{A Yangian double for the AdS/CFT classical r-matrix}",
    eprint = "0706.0884",
    archivePrefix = "arXiv",
    primaryClass = "hep-th",
    reportNumber = "MIT-CTP-3843",
    doi = "10.1088/1126-6708/2007/06/083",
    journal = "JHEP",
    volume = "06",
    pages = "083",
    year = "2007"
}

@article{Beisert:2010kk,
  author = "Beisert, Niklas",
  title = "The Classical Trigonometric r-Matrix for the Quantum-Deformed Hubbard Chain",
  eprint = "1002.1097",
  archivePrefix = "arXiv",
  primaryClass = "math-ph",
  reportNumber = "AEI-2010-016",
  doi = "10.1088/1751-8113/44/26/265202",
  journal = "J. Phys. A",
  volume = "44",
  pages = "265202",
  year = "2011"
}

@article{Zoubos:2010kh,
    author = "Zoubos, Konstantinos",
    title = "{Review of AdS/CFT Integrability, Chapter IV.2: Deformations, Orbifolds and Open Boundaries}",
    eprint = "1012.3998",
    archivePrefix = "arXiv",
    primaryClass = "hep-th",
    doi = "10.1007/s11005-011-0515-8",
    journal = "Lett. Math. Phys.",
    volume = "99",
    pages = "375--400",
    year = "2012"
}

@article{Berenstein:2005vf,
    author = "Berenstein, David and Vazquez, Samuel E.",
    title = "{Integrable open spin chains from giant gravitons}",
    eprint = "hep-th/0501078",
    archivePrefix = "arXiv",
    reportNumber = "NSF-KITP-05-01",
    doi = "10.1088/1126-6708/2005/06/059",
    journal = "JHEP",
    volume = "06",
    pages = "059",
    year = "2005"
}

@article{MacKay:2010ey,
    author = "MacKay, Niall and Regelskis, Vidas",
    title = "{Yangian symmetry of the Y=0 maximal giant graviton}",
    eprint = "1010.3761",
    archivePrefix = "arXiv",
    primaryClass = "hep-th",
    doi = "10.1007/JHEP12(2010)076",
    journal = "JHEP",
    volume = "12",
    pages = "076",
    year = "2010"
}

@article{MacKay:2011zs,
    author = "MacKay, Niall and Regelskis, Vidas",
    title = "{Achiral boundaries and the twisted Yangian of the D5-brane}",
    eprint = "1105.4128",
    archivePrefix = "arXiv",
    primaryClass = "hep-th",
    doi = "10.1007/JHEP08(2011)019",
    journal = "JHEP",
    volume = "08",
    pages = "019",
    year = "2011"
}

@article{deLeeuw:2012fy,
    author = "de Leeuw, Marius and Regelskis, Vidas",
    title = "{Integrable boundaries in AdS/CFT: revisiting the Z=0 giant graviton and D7-brane}",
    eprint = "1206.4704",
    archivePrefix = "arXiv",
    primaryClass = "hep-th",
    doi = "10.1007/JHEP03(2013)030",
    journal = "JHEP",
    volume = "03",
    pages = "030",
    year = "2013"
}

@article{Palla:2011eu,
    author = "Palla, Laszlo",
    title = "{Yangian symmetry of boundary scattering in AdS/CFT and the explicit form of bound state reflection matrices}",
    eprint = "1102.0122",
    archivePrefix = "arXiv",
    primaryClass = "hep-th",
    reportNumber = "ITP-BUDAPEST-REPORT-650",
    doi = "10.1007/JHEP03(2011)110",
    journal = "JHEP",
    volume = "03",
    pages = "110",
    year = "2011"
}

@article{Ciavarella:2011jca,
    author = "Ciavarella, Andrew Michael",
    title = "{Giant Magnons and Giant Gravitons}",
    note = "PhD thesis, Durham U., England",
    year = "2011"
}

@article{Galleas:2009ye,
    author = "Galleas, W.",
    title = "{The Bethe Ansatz Equations for Reflecting Magnons}",
    eprint = "0902.1681",
    archivePrefix = "arXiv",
    primaryClass = "hep-th",
    reportNumber = "AEI-2009-012",
    doi = "10.1016/j.nuclphysb.2009.04.024",
    journal = "Nucl. Phys. B",
    volume = "820",
    pages = "664--681",
    year = "2009"
}

@article{Nepomechie:2009zi,
    author = "Nepomechie, Rafael I.",
    title = "{Bethe ansatz equations for open spin chains from giant gravitons}",
    eprint = "0903.1646",
    archivePrefix = "arXiv",
    primaryClass = "hep-th",
    reportNumber = "UMTG-263",
    doi = "10.1088/1126-6708/2009/05/100",
    journal = "JHEP",
    volume = "05",
    pages = "100",
    year = "2009"
}

@article{Murgan:2008zu,
    author = "Murgan, Rajan and Nepomechie, Rafael I.",
    title = "{q-deformed su(2$/$2) boundary S-matrices via the ZF algebra}",
    eprint = "0805.3142",
    archivePrefix = "arXiv",
    primaryClass = "hep-th",
    reportNumber = "UMTG-257",
    doi = "10.1088/1126-6708/2008/06/096",
    journal = "JHEP",
    volume = "06",
    pages = "096",
    year = "2008"
}

@article{Ahn:2007bq,
    author = "Ahn, Changrim and Bak, Dongsu and Rey, Soo-Jong",
    title = "{Reflecting Magnon Bound States}",
    eprint = "0712.4144",
    archivePrefix = "arXiv",
    primaryClass = "hep-th",
    reportNumber = "SNUST-071201, UOSTP-07107",
    doi = "10.1088/1126-6708/2008/04/050",
    journal = "JHEP",
    volume = "04",
    pages = "050",
    year = "2008"
}

@article{Ahn:2008df,
    author = "Ahn, Changrim and Nepomechie, Rafael I.",
    title = "{The Zamolodchikov-Faddeev algebra for open strings attached to giant gravitons}",
    eprint = "0804.4036",
    archivePrefix = "arXiv",
    primaryClass = "hep-th",
    reportNumber = "UMTG-256",
    doi = "10.1088/1126-6708/2008/05/059",
    journal = "JHEP",
    volume = "05",
    pages = "059",
    year = "2008"
}

@article{Alcaraz:1987uk,
    author = "Alcaraz, F. C. and Barber, M. N. and Batchelor, M. T. and Baxter, R. J. and Quispel, G. R. W.",
    title = "{Surface Exponents of the Quantum XXZ, Ashkin-Teller and Potts Models}",
    reportNumber = "ANU-RR-7-1987",
    doi = "10.1088/0305-4470/20/18/038",
    journal = "J. Phys. A",
    volume = "20",
    pages = "6397",
    year = "1987"
}

@Book{Chari:1994pz,
     author    = "Chari, V. and Pressley, A.",
     title     = "A guide to quantum groups",
     address   = "Cambridge, UK",
     year      = "1994",
     publisher = "Cambridge University Press"
}

@book{Gomez:1996az,
    author = "Gómez, C. and Sierra, G. and Ruiz-Altaba, M.",
    title = "{Quantum groups in two-dimensional physics}",
    doi = "10.1017/CBO9780511628825",
    isbn = "978-0-511-88509-9, 978-0-521-02004-6, 978-0-521-46065-1",
    publisher = "Cambridge University Press",
    series = "Cambridge Monographs on Mathematical Physics",
    month = "4",
    year = "2011"
}

\end{bibtex}
\fi

\bibliographystyle{nb}
\bibliography{\jobname}

\end{document}